\documentclass{aa}  
\usepackage{graphicx}
\usepackage{txfonts}
\usepackage{longtable}
\begin{document} 

   \title{Stellar activity and rotation of the planet host Kepler-17 \\
   from long-term space-borne photometry}
   \authorrunning{Lanza et al.}
    \titlerunning{Kepler-17 activity and rotation}
   \subtitle{}
\author{
A.~F.~Lanza,\inst{1}
Y.~Netto,\inst{1,2} 
A.~S.~Bonomo,\inst{3} 
H.~Parviainen,\inst{4}  
A.~Valio,\inst{2}
S.~Aigrain\inst{5}  
}
\institute{INAF -- Osservatorio Astrofisico di Catania, Via S.~Sofia 78,  I-95123 Catania, Italy\\
              \email{antonino.lanza@inaf.it} 
\and CRAAM, Mackenzie Presbyterian University, Rua da Consolacao, 896, Sao Paulo, Brazil
\and  INAF -- Osservatorio Astrofisico di Torino, Via Osservatorio 20, I-10025, Pino Torinese, Italy  
\and Instituto de Astrof\'{\i}sica de Canarias, V\'{\i}a L\'actea s/n, E-38205 La Laguna, Spain
\and Sub-department of Astrophysics, Department of Physics, University of Oxford, Oxford OX1 3RH, UK}

   \date{Received ... ; accepted ...}

\abstract{The study of young Sun-like stars is of fundamental importance to understand the magnetic activity and rotational evolution of the Sun. Space-borne photometry by the Kepler telescope provides unprecedented datasets to investigate these phenomena in Sun-like stars.}{We present a new analysis of the entire Kepler photometric time series of the moderately young Sun-like star Kepler-17 that is accompanied by a transiting hot Jupiter.}{We applied a maximum-entropy spot model to the long-cadence out-of-transit photometry of the target to derive maps of the starspot filling factor versus the longitude and the time. These maps are compared to the spots occulted during transits to validate  our reconstruction and derive information on the latitudes of the starspots.}{We find two main active longitudes on the photosphere of Kepler-17, one of which has a lifetime of at least $\sim 1400$~days although with a varying level of activity.  The latitudinal differential rotation is of solar type, that is, with the equator rotating faster than the poles. We estimate a minimum relative amplitude $\Delta \Omega/ \Omega$ between $\sim 0.08 \pm 0.05$ and $0.14 \pm 0.05$, our determination being affected by the finite lifetime of individual starspots and depending on the adopted spot model parameters. We find marginal evidence of a short-term intermittent activity cycle of $\sim 48$~days and an indication of a longer cycle of $400-600$~days characterized by an equatorward migration of the mean latitude of the spots as in the Sun. The rotation of Kepler-17 is likely to be significantly affected by the tides raised by its massive close-by planet.}{We confirm  the reliability of maximum-entropy spot models to map starspots in young active stars and characterize the activity and differential rotation of this young Sun-like planetary host. }
   \keywords{stars: activity -- stars: rotation -- stars: late-type -- starspots -- planetary systems -- techniques: photometry} 
  \maketitle
%

\section{Introduction}
The interaction of convection and rotation produces differential rotation and magnetic fields in the Sun. On timescales of billions of years, magnetic fields affect the rotation of our star by the angular momentum loss associated with  its magnetized stellar wind. A better understanding of these complex processes and their interconnections can be obtained by comparing the Sun with other late-type stars, in particular with younger Sun-like stars that show a higher level of magnetic activity \citep[cf.][]{BrunBrowning17}. Space-borne telescopes, such as {\it CoRoT} \citep{Auvergneetal09} or {\it Kepler} \citep{Kochetal10}, allow us to monitor late-type stars photometrically in the optical passband with a relative accuracy down to $10^{-5}-10^{-4}$ with integration times from minutes to hours for time intervals up to $3-4$ years, thus providing unprecedented datasets to study their magnetic activity. 

Magnetic fields, produced in the stellar interior by hydromagnetic dynamos, emerge into the photosphere where they modify the transport of energy and momentum giving rise to cooler and hotter patches, called starspots and faculae, respectively \citep[e.g.,][]{Gondoin08,Strassmeier09}, that modulate the optical flux integrated over the stellar disc owing to their intrinsic evolution and the rotation of the star. This flux modulation can be modelled  to extract information on the locations of the surface brightness inhomogeneities and  their evolution. {Recent works have explored the possibility offered by starspots as tracers of stellar rotation and differential rotation \citep[e.g.,][]{Mosseretal09,Walkowiczetal13,Santosetal17}, while a general review of the different approaches to spot modelling can be found in, e.g., \citet{Lanza16} and a comparison between spot models and sunspot group observations in \citet{Lanzaetal07}}. Models are better constrained when the inclination of the stellar spin axis to the line of sight is known such as in the case of stars that have a transiting planet \citep[e.g.,][]{Winnetal05,Nutzmanetal11}. 

Among the stars with a transiting hot Jupiter, Kepler-17 is one of the targets with more extended and precise transit observations \citep{Muelleretal13} that allowed to map the spots occulted during the transits \citep{EstrelaValio16,Valioetal17} and  constrain the inclination of the stellar spin axis \citep{Desertetal11}. Moreover, it is a young star of G2V spectral type making it an ideal candidate for solar-stellar connection studies. It has an estimated age of $\la 1.8$~Gyr, a mean rotation period of $\sim 12$~days, and is accompanied by a planet with a mass of $2.47 \pm 0.10$ Jupiter masses, a radius of $1.33 \pm 0.04$ Jupiter radii, and an orbital period $P_{\rm orb} = 1.48571$~days \citep{Bonomoetal12}. 

\citet{BonomoLanza12} analysed  $\sim 500$~days of public Kepler data available at that time; now the availability of the latest Kepler data release with a high-precision photometric time series extending for $\sim 1500$~days calls for a new modelling of this star to study its activity and rotation using starspots as tracers. Moreover, we compare the longitudes of the spots mapped from the out-of-transit photometry with those of the spots occulted during transits, providing an independent confirmation of our results and giving constraints on the spot latitudes that cannot be obtained with alternative methods in the case of Kepler-17. In such a way, we  investigate the differential rotation of our target, the phenomenology of its active longitudes, and its activity cycles. 

The presence of a close-by giant planet affects the properties of Kepler-17, notably its  rotation {that is used to estimate its age by applying the gyrochronology technique \citep{Barnes07,Barnes10}. We look for features associated with star-planet interaction in the photometric time series and investigate the tidal evolution of the system to clarify the difference in the evolution of the stellar rotation with respect to a single star such as our Sun.  Moreover, the peculiar evolution of the rotation of Kepler-17 affects the flux of the stellar high-energy radiation experienced by the planet during its lifetime.  }

\section{Observations}
\label{observations}
The Kepler 95-cm telescope was designed to continuously look at a fixed field in the Cygnus constellation to detect planetary transits. Every three months, the spacecraft is rolled by $90^{\circ}$ about its line of sight to keep its solar panels pointing towards the Sun. Each of these  periods is called a quarter in the Kepler jargon. Because of the rotation of the focal plane, each target falls on  different CCDs during different quarters, so the observations must be reduced quarter-by-quarter. The usual cadence of Kepler observation is 1765.5~s (long cadence), although for a subset of interesting targets, such as those showing transits, the cadence is reduced to 58.5~s (short cadence). We shall analyse photometry acquired in long cadence because we are interested in the activity and rotation of Kepler-17 both of which have timescales of the order of several days, while short cadence has been used to observe planetary transits and detect spots occulted during transits \citep{EstrelaValio16,Valioetal17}. 

In the mission archive\footnote{https://archive.stsci.edu/kepler/} there are the time series of the electron counts in the individual CCD pixels within a pre-defined area around the image of each target  in the focal plane and two light curves. The first is obtained by summing the flux falling within a subset of the pixels included in the target pixel files; a correction for the background flux is also applied. This is called the Simple Aperture Photometry (hereinafter SAP) light curve. The second time series is obtained by additional processing to remove instrumental and systematic effects and  is called the Pre-search Data Conditioning (hereafter PDC) light curve \citep{Stumpeetal12,Stumpeetal14}.  

We downloaded from the mission archive all the long-cadence SAP time series of the latest data release (Data Release 25) of Kepler-17 covering 14 quarters out of a total of 18. This time series is affected by outliers and systematic instrumental effects that have been corrected in the PDC time series. However, the PDC time series shows a significant reduction of the amplitude of intrinsic stellar variability and sometimes a distortion of the modulations on time scales longer than a few days because it has been designed to detect planetary transits, not to preserve the intrinsic stellar variability \citep[see][]{Gillilandetal15}. These effects are particularly relevant for Kepler-17 because it shows a large light curve amplitude of the order of 0.05 mag with typical modulation timescales $\geq 10$ days. For these reasons, we decided not to use the PDC time series in the present analysis and derived two light curves starting from the SAP time series. 

The first was derived by means of a procedure called ARC2 introduced by \citet{Aigrainetal17} that warrants a better preservation of the intrinsic stellar variability while removing discontinuities, outliers, and instrumental effects by making use of the Co-trending Basis Vectors (CBVs) computed by the PDC pipeline. CBVs describe instrumental effects for each target and are based on the systematics observed in targets that are close on the CCD to the given target and are similar in flux. They also take into account instrumental effects by making use of the spacecraft telemetry information. Therefore, they provide the best available description of the systematics affecting the time series of a given target. Up to eight CBVs are used by the PDC pipeline to perform its correction often leading to an overcorrection of the instrinsic stellar variability on timescales much longer than those characteristics of planetary transits and to an injection of additional noise on those timescales. On the other hand, the ARC2 pipeline of \citet{Aigrainetal17} was designed to preserve the intrinsic target variability and reduce the injected noise in the correction process as much as possible and is particularly valuable in the case of bright and remarkably variable stars such as Kepler-17 whose light curves are modified by the PDC pipeline. 

{In our application, we started from the SAP time series considering only the datapoints with a SAP\_QUALITY flag equal to zero, that is without any detected problem during their acquisition \citep{Jenkinsetal16}. The ARC2 pipeline includes a Bayesian method to find the best weights to calculate the linear combination of the CBVs to correct each  quarter time series \citep[see Sect. 3.1 in][]{Aigrainetal17} as well as a criterion to select the optimal number of CBVs to be used. The latter is based on balancing the reduction of the normalized light curve amplitude, that comes from removing systematics by adding successive CBVs, against the increase of the short-term noise resulting from the same operation \citep[see Sect. 3.2 of][for a quantitative description of the criterion]{Aigrainetal17}. }

\begin{figure*}
\hspace*{-7mm}
 \centering{
 \includegraphics[width=10cm,height=20cm,angle=270]{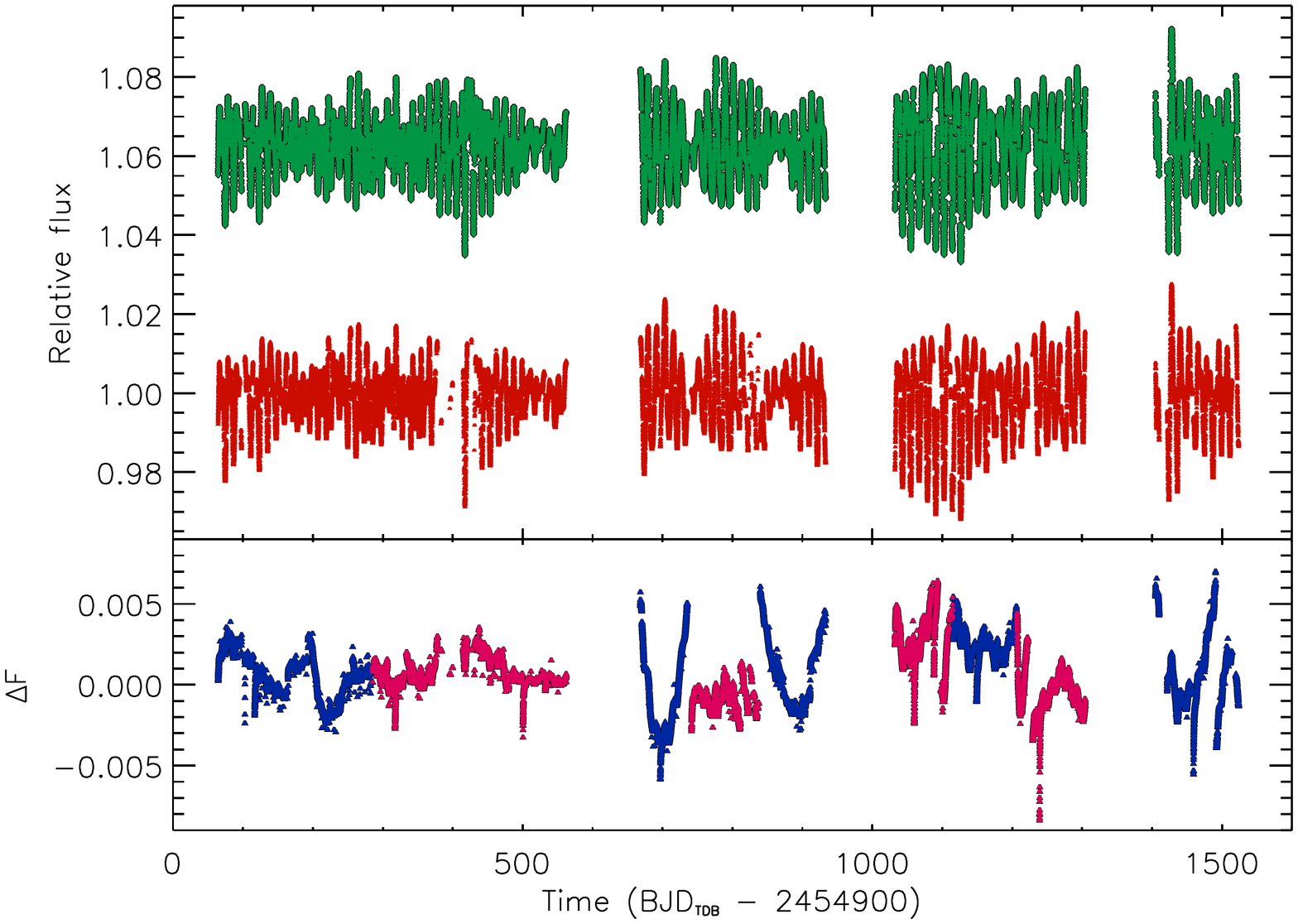}} 
   \caption{Top panel: The light curve of Kepler-17 as obtained by applying the de-trending as in \citet{BonomoLanza12} (in green) and the light curve as provided by the ARC2 pipeline (in red).  
  Each light curve is normalized to the median flux within each quarter after removing the planetary transits. 
   The green datapoints have been shifted upward by 0.0625 for clarity. Bottom panel: Difference between the two light curves. {The quarters where only the first CBV was applied to obtain the ARC2 light curve are indicated with blue datapoints, while the quarters where the first two CBVs  have been applied are indicated in pink (see the text). } }
              \label{fig1}%
\end{figure*}

We removed the planetary transits from the light curve provided by ARC2  using the ephemeris of \citet{Muelleretal13}, discarding the datapoints before 0.05 and after 0.05 in phase of the first and the fourth contacts, respectively. Then each quarter was normalized to its median value and residual outliers were removed by applying a 3-$\sigma$ clipping to the residuals obtained by subtracting a smoothed version of the light curve obtained with a boxcar filter with a width of 294 minutes, that is ten consecutive datapoints.  A total of 340 datapoints were flagged as outliers and discarded. The final lightcurve consists of 40653 datapoints and ranges from BJD$_{\rm TDB}$\footnote{{We measure the time as Barycentric Julian Date (BJD) in the Barycentric Dynamical Time (TDB) at the mid-point of each photometric exposure  \citep[see][]{Eastmanetal10}.  The difference with respect to other definitions of BJD is always smaller than 1~minute that is negligible for our purposes.}} 2454964.512 to 2456423.980 covering a total of 1459.469 days with four main gaps (see Fig.~\ref{fig1}, top panel red plot). The median of the error of the data points is $2.54 \times 10^{-4}$ in relative flux units. 

To allow a straightforward comparison with the results previously obtained by \citet{BonomoLanza12}, a second light curve was derived from the SAP time series following their approach. {In this case, we choose to consider all the datapoints with a finite flux value, not limiting ourselves to those with a SAP\_QUALITY flag equal to zero, because a different flag does not indicate in general that the datapoint is unusable as described in the Kepler Archive Manual Sect.~2.3.1.1. In brief, the procedure used to obtain the analysed light curve was as follows.} Data were corrected quarter-by-quarter by {discarding  the steep variations after the safe modes or data download links; then we  removed the points in transits as in the case of the ARC2 light curve}. Next, long-term trends of clear instrumental origin were removed by fitting a parabola. Finally, each quarter was normalized to its median value and the 3-$\sigma$ clipping procedure used for the ARC2 light curve was applied to flag and remove outliers. In such a way, a total of 1219 datapoints were discarded. The final light curve consists of 47376 datapoints ranging from BJD$_{\rm TDB}$ 2454964.512 to 2456424.001 for a total of 1459.489 days with three main gaps. The number of datapoints is greater than in the ARC2 light curve because there are points covering the first gap in the ARC2 light curve and some others covering some small gaps in other intervals of the light curve (see Fig.~\ref{fig1}, top panel green plot) {thanks to our less strict selection criterion including datapoints with non-zero SAP\_QUALITY flag.} The median of the error of the data points is $2.22 \times 10^{-4}$ in relative flux units. 

The difference between the two light curves is plotted in the bottom panel of Fig.~\ref{fig1}. The flux values of the light curve obtained with the approach of \citet{BonomoLanza12} were linearly interpolated at the epochs of the ARC2 time series to compute the flux difference. {Such an interpolation was in order because there were 492 datapoints of the ARC2 light curve that did not have a corresponding datapoint at the same time in the other light curve owing to the different criteria adopted to reject points affected by systematics close to data gaps or steep variations (automated removal in the case of ARC2, manual removal in the other case).} The correction computed by the ARC2 pipeline makes use of only the first CBV for eight quarters and of the first two CBVs for the remaining six quarters, providing a better preservation of the intrinsic stellar variability than the PDC light curve. This is confirmed by the small difference with the light curve computed with the approach of \citet{BonomoLanza12} for which the long-term trends inside each quarter were corrected by a simple parabolic fit and  the steep variations detected by eye simply eliminated. This produces a more flat light curve, i.e., showing less modulation of its mean level on timescales comparable with a quarter or longer. {The parabolic shape of the difference between the two light curves in some quarters (see Fig.~\ref{fig1}, lower panel) corresponds to quarters where only the first CBV was used for the correction of the ARC2 light curve. }

\section{Methods}
\label{methods}
The spot modelling approach applied in the present study is the same already introduced in Sect. 3 of \citet{BonomoLanza12} to which we refer the reader for details. In brief, the surface of the star is subdivided into 200 surface elements that contain unperturbed photosphere, dark spots, and solar-like faculae. The specific intensity of the unperturbed photosphere in the Kepler passband is assumed to vary according to a quadratic limb-darkening law:
\begin{equation}
I(\mu) = I_{0} (a_{\rm p} + b_{\rm p} \mu + c_{\rm p} \mu^{2}), 
\label{limbdark}
\end{equation}
where $I_{0}$ is the specific intensity at the centre of the disc, $\mu = \cos \theta$ with $\theta$ being the angle between the local surface normal and the line of sight, and $a_{\rm p}$, $b_{\rm p}$, and $c_{\rm p}$ are the limb-darkening coefficients in the Kepler passband. 

The dark spots are assumed to have a fixed contrast $c_{\rm s} \equiv I_{\rm spot}(\mu)/I(\mu)$ in the Kepler passband, where $I_{\rm spot}$ is the specific intensity in the spotted photosphere. The fraction of a surface element covered by dark spots is given by its filling factor $f$. 
The faculae are assumed to have a fixed contrast $c_{\rm f} =1.115$ at the limb that varies linearly with $\mu$ becoming unity (no flux perturbation) at the centre of the disc. In this way, they mimic the contrast behaviour of solar photospheric faculae, at least in a rough and average sense. The ratio $Q$ of their area to that of the dark spots is fixed, so that their filling factor is $Qf$. {In our model, $Q$ always appears in combination with $c_{\rm f}$ in the product $c_{\rm f}Q$. Therefore, it is sufficient to vary $Q$ to change the contribution of the faculae in our model \citep[cf.][]{Lanzaetal07,Lanza16}.}

This model is fitted to a segment of the light curve of duration $\Delta t_{\rm f}$ (see Sect.~\ref{parameters}) by varying the filling factors of the individual surface elements that can be represented as a 200-element vector $\vec f$. Therefore, the model has 200 free parameters and suffers from  non-uniqueness and instability due to the effect of photometric noise. To select a unique and stable solution, we apply a maximum entropy regularization by minimizing a functional $Z$ that is a linear combination of the $\chi^{2}$ and of a suitable entropy function $S$: 
\begin{equation}
Z = \chi^{2}(\vec f) - \lambda S(\vec f),
\end{equation}
where $\lambda > 0$ is a Lagrangian multiplier that controls the relative weights given to the $\chi^{2}$ minimization and the configuration entropy of the surface map $S$ in the solution.  The expression of $S$ is given in Eq.~(5) of \citet{BonomoLanza12} and it is maximal  when the star is unspotted, that is all the elements of the vector $\vec f$ are zero. In other words, the maximum entropy (hereafter ME) criterion selects the solution with the minimum spotted area compatible with a given $\chi^{2}$ value of the best fit to the light curve.   When the Lagrangian multiplier $\lambda = 0$, we obtain the solution corresponding to the minimum $\chi^{2}$ that is unstable. By increasing $\lambda$, we obtain a unique and stable solution at the price of increasing the value of the $\chi^{2}$. An additional effect is that of making the residuals between the model and the light curve biased towards negative values because we reduce the spot filling factors by introducing the entropy term \citep[see][for more details]{Lanza16}. 

The information on the latitude of the spots is lacking in our maximum-entropy maps because the inclination of the stellar spin axis is very close to $90^{\circ}$ (cf. Sect.~\ref{parameters}) that makes the transit time of each  feature independent of its latitude. Therefore, we shall limit ourselves to map the distribution of the filling factor versus the longitude.  

The optimal value of the Lagrangian multiplier $\lambda$ is obtained by imposing that the mean $\mu_{\rm reg}$ of the residuals between the regularized model and the light curve verifies the relationship \citep{BonomoLanza12,Lanza16}:
\begin{equation}
| \mu_{\rm reg} | = \frac{\sigma_{0}}{\sqrt{N}},  
\end{equation}
where $\sigma_{0}$ is the standard deviation of the residuals of the unregularized model, that is that computed with $\lambda =0$, and $N$ the number of datapoints in the fitted light curve interval of duration $\Delta t_{\rm f}$.

The optimal value of $\Delta t_{\rm f}$ is not known a priori and must be determined with an analysis of the light curve itself because it is related to the lifetimes of the active regions in a given star. We shall adopt a unique value of $\Delta t_{\rm f}$ for the entire light curve of Kepler-17  because the ratio $\Delta t_{\rm f}/P_{\rm rot}$, where $P_{\rm rot}$ is the stellar rotation period, rules the sensitivity of the spot modelling to active regions located at different longitudes as discussed by \citet{Lanzaetal07}. 

The optimal value of the facular-to-spotted area ratio $Q$ is also derived from the light curve best fit. To find the best values of $\Delta t_{\rm f}$ and $Q$, we use a simple spot model consisting of three active regions and a varying background level that was  introduced in the case of the Sun \citep[cf. Sect.~3 of][]{Lanzaetal03} and previously applied by \citet{BonomoLanza12} to Kepler-17. It  has a much smaller number of free parameters than the maximum-entropy model, allowing us a faster exploration of the $\Delta t_{\rm f}$-$Q$ parameter space to look for the combination that minimizes the $\chi^{2}$ of the entire light curve. 

{In principle, $\Delta t_{\rm f}$ and $Q$ are not  parameters of the same kind because the former is the extension of the individually fitted time intervals, while the latter is a model parameter. However, $\Delta t_{\rm f}$ affects the determination of  $Q$ because $Q$ is constrained by the different facular contrasts between the centre of the disc and the limb. In other words, the determination of $Q$ is affected by the duration of the intervals during which the model active regions are close to the centre or to the limb which in turn depend on  $\Delta t_{\rm f}$. For example, too a short $\Delta t_{\rm f}$ does not allow the model active regions to move all along their chords across the stellar disc, so they cannot span the full range between the centre of the disc and the limb and provide the best constraint on $Q$. In this case, the dependence of the $\chi^{2}$ of the entire light curve on $Q$ is characterized by random oscillations. For this reason, we optimize $Q$ for several fixed values of $\Delta t_{\rm f}$ considering the range where the total $\chi^{2}$ of the 3-spot model depends regularly on $Q$ and look for the minimum $\chi^{2}$  in the $\Delta t_{\rm f}$-$Q$ space. }

\section{Stellar parameters}
\label{parameters}
The basic stellar parameters, that is mass, radius, effective temperature $T_{\rm eff}$, and surface gravity $\log g$, are taken from \citet{Bonomoetal12} and are the same adopted by \citet{BonomoLanza12}. They do not directly enter into our geometric spot model, except for the computation of the ratio $\epsilon_{\rm rot}$ between the polar and the equatorial axes of the ellipsoid used to represent the surface of the star. Its value is obtained by a simple Roche model assuming rigid rotation with a period of $P_{\rm rot} = 12.01$ days as in \citet{BonomoLanza12}. {The gravity darkening effect associated with $\epsilon_{\rm rot} \sim 4.7 \times 10^{-5}$ is of the order of a few times $10^{-6}$~mag, thus it can be neglected in our model. }

The best fit to the transits of Kepler-17b can be used to extract information on the inclination $i$ of the stellar spin axis to the line of sight and on the quadratic limb-darkening coefficients. \citet{Muelleretal13} provided a refined analysis of the first seven quarters of Kepler short-cadence observations adopting the stellar parameters of \citet{Bonomoetal12} and compared the limb-darkening parameters derived from the transit fit with those given by model atmospheres.  The precision of the limb-darkening parameters derived from the fitting of the average transit profile is remarkably high because Kepler-17 has a photometry with a high signal-to-noise ratio. Nevertheless, the theoretical linear coefficient is significantly different from that derived from the transit fitting. The discrepancy is more clearly  evident in the recent analysis by \citet{Maxted18} (cf. his Fig.~4) who considered all the short-cadence quarters available in Kepler Data Release 25 and suggested that the remarkable activity of Kepler-17 is responsible for it. In other words, the  active regions on the stellar surface, the most part of which is not resolved, even by methods applied to detect spot occultations during transits, make the limb-darkening profile of Kepler-17 significantly different from that computed from model atmospheres. {In our spot modelling, we shall adopt the limb-darkening coefficients as derived from the transit fitting, thus modelling the out-of-transit light curve in a way consistent with that adopted for the occulted spots. However, we shall investigate the impact of this choice by computing additional models with the theoretical limb-darkening coefficients to show that our main results are not critically dependent on those coefficients (cf. Sect.~\ref{results} and Appendix~\ref{app1}). Specifically, we shall adopt the coefficients as derived by \citet{Muelleretal13}  from a model atmosphere with $T_{\rm eff} = 5787$~K, $\log g = 4.45$, and solar metallicity (cf. their Sect.~5.1). } 

In consideration of his use of the full dataset of the latest Kepler Data Release, we assume the transit fit by \citet{Maxted18} as the best for our purposes and derive the quadratic limb-darkening coefficients from his model. Specifically, given that \citet{Maxted18} adopted a non-polynomial limb-darkening law of the form $I(\mu)/I_{0} = 1 - c(1-\mu^{\alpha})$ with free parameters $c$ and $\alpha$, we fitted our quadratic law in Eq.~(\ref{limbdark}) to $10^{5}$ realizations of this equation with the values of the coefficients $c$ and $\alpha$ drawn from their a posteriori distributions. The quadratic fit is good, except for a relative deviation of $\sim 2-3$ percent close to the limb ($\mu > 0.8$), that can be neglected because the photometric effect of starspots close to the limb is negligible owing to the strong reduction of their projected area by foreshortening. In such a way, we obtain the limb-darkening coefficients in Table~\ref{model_param} as the medians of their a posteriori distributions. {Note that we cannot adopt the same functional form of the limb darkening as in \citet{Maxted18} because our computer code should  have been completely rewritten given that it makes use of the quadratic form of the limb-darkening law to speed up calculations.}

The inclination of the stellar spin axis is assumed equal to that of the orbital plane as derived from the transit fitting as in \citet{BonomoLanza12}. Adopting the parameters of the model of \citet{Maxted18}, we obtain the median value listed in Table~\ref{model_param} that is also compatible with that of \citet{Muelleretal13} and with the estimate coming from the rotation spectral line broadening $v \sin i$, $P_{\rm rot}$ and the estimated stellar radius \citep{BonomoLanza12}. The mean rotation period $P_{\rm rot}$ is adopted as in \citet{BonomoLanza12} for a straightforward comparison with their results.

The contrast of the dark spots was assumed to be equal to that of sunspot groups, that is $c_{\rm s} = 0.677$, by \citet{BonomoLanza12}. However, the recent work by \citet{Valioetal17} provides a direct measure based on the modelling of starspots occulted during transits. Their modelling gives the mean value $c_{\rm s} = 0.55 \pm 0.17$  that is adopted in the present analysis. {However, in view of the large differences in $c_{\rm s}$ among the different spots modelled by \citet{Valioetal17}, in Sect.~\ref{results} and in Appendix~\ref{app1}, we explore the effects of the variation of the spot contrast on our results by computing regularized models with the two extreme values $c_{\rm s} = 0.38$ and $0.72$.}
\begin{table}
\caption{Parameters adopted for the modelling of the light curves of Kepler-17.}
\begin{center}
\begin{tabular}{lcc}
\hline
\hline
Parameter & Value & Ref. \\
\hline
Star mass ($M_{\odot}$)     & 1.16 & B12 \\
Star radius($R_{\odot}$)   & 1.05 & B12 \\
$T_{\rm eff}$ (K) & 5780  & B12 \\
$\log g $ (cm\,s$^{-2}$) & 4.53 & B12 \\
$a_{\rm p}$ & 0.581 & L19 \\
$b_{\rm p}$ & 0.340 & L19 \\
$c_{\rm p}$ & 0.079 & L19 \\
$P_{\rm rot}$ (days) & 12.01 & BL12\\
$\epsilon_{\rm rot}$ & $4.66 \times 10^{-5}$ & BL12 \\
$i$ (deg) & 89.88 & M18 \\
$c_{\rm s}$ & 0.550 & V17\\
$c_{\rm f}$  & 0.115 & BL12 \\
$Q$ & 2.4 & L19 \\
$\Delta t_{\rm f}$ (days) & 8.733 & L19\\
 \hline
\noindent
\end{tabular}
\end{center}
{References.} BL12: \citet{BonomoLanza12}; B12: \citet{Bonomoetal12}; L19: present study; M18: \citet{Maxted18}; V17: \citet{Valioetal17}. 
\label{model_param}
\end{table}

\begin{figure}
\hspace*{-1cm}
 \centering{
 \includegraphics[width=8cm,height=10cm,angle=90]{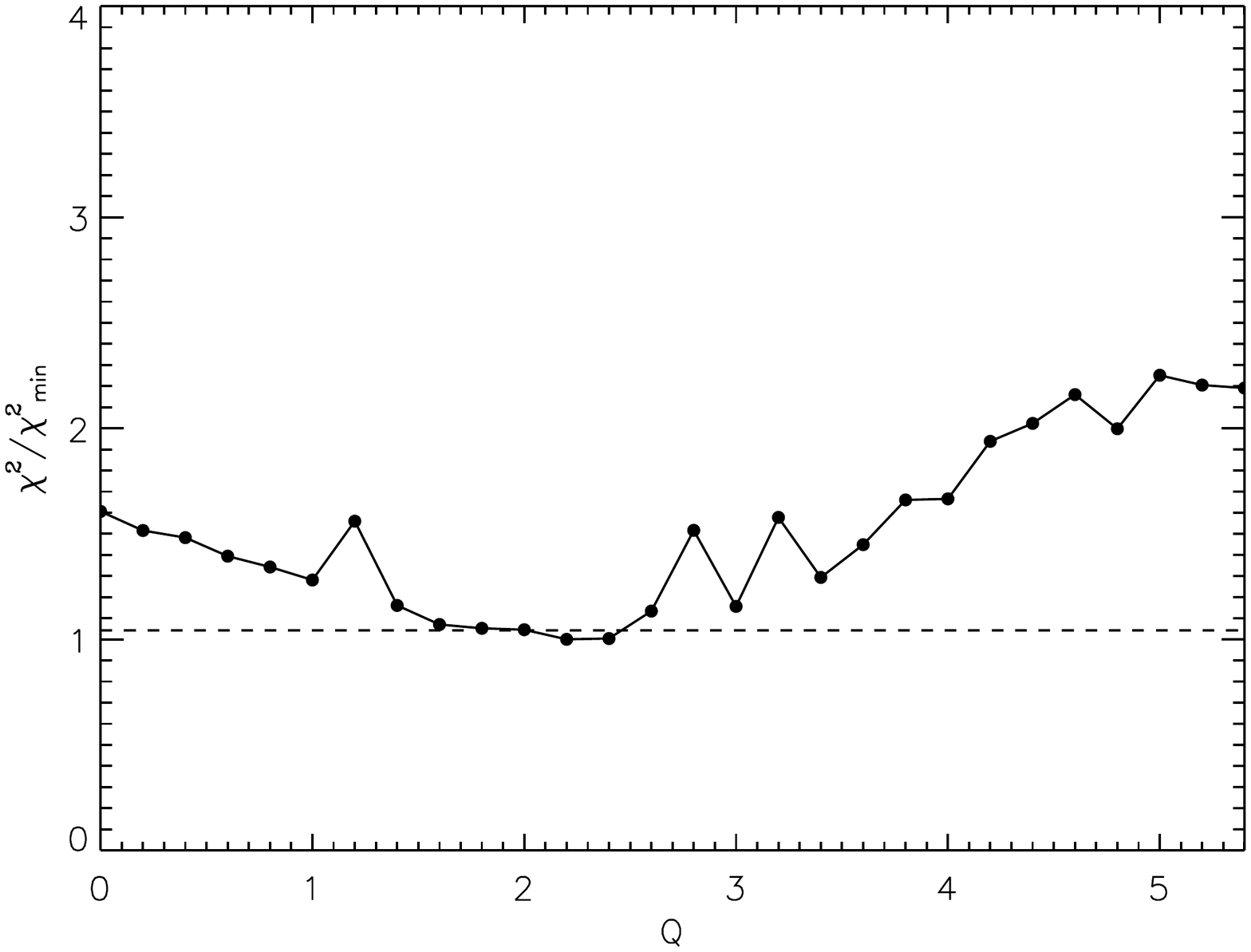}} 
    \caption{Ratio of the $\chi^{2}$ of the best fit to the entire ARC2 light curve
  to its minimum value vs. the parameter $Q$, i.e., the ratio
of the area of the faculae to that of the dark spots in active regions.
The horizontal dashed line indicates the 95 percent confidence level for $\chi^{2}/\chi^{2}_{\rm min}$ determining the interval of acceptable $Q$ values. }
              \label{fig2}%
\end{figure}

\begin{figure}
\hspace*{-1cm}
 \centering{
 \includegraphics[width=8cm,height=10cm,angle=90]{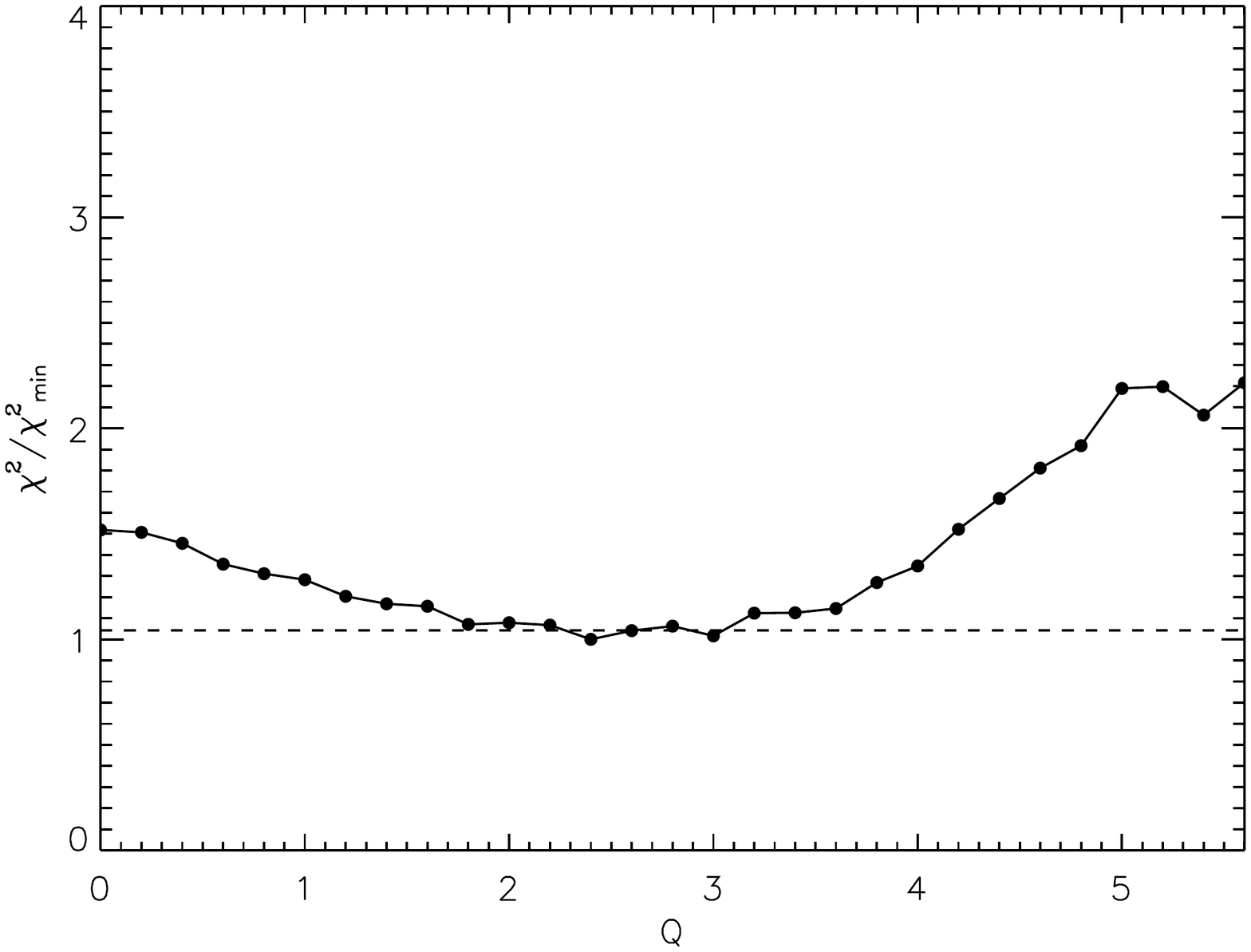}} 
 \caption{Same as Fig.~\ref{fig2} for the best fit to the light curve obtained with the de-trending of \citet{BonomoLanza12}. }
              \label{fig3}%
\end{figure}

The facular-to-spotted area ratio $Q$ was derived by using the simplified three-spot model to fit the entire light curve of Kepler-17 by selecting the value that gives the minimum total chi square. The plot of the ratio between the $\chi^{2}$ and its minimum $\chi^{2}_{\rm min}$ vs. $Q$ is shown in Fig.~\ref{fig2} for the ARC2 light curve and in Fig.~\ref{fig3} for the light curve de-trended as in \citet{BonomoLanza12} assuming $\Delta t_{\rm f} = 8.733$~days. {The ratio of the $\chi^{2}$ to its minimum is statistically distributed according to the Fischer-Snedecor statistics $F$ as \citep[cf.][Sect.~VI]{Lamptonetal76}:
\begin{equation}
\frac{\chi^{2}}{\chi^{2}_{\rm min}} \sim 1 + \frac{p}{N-p} F(p, N-p),
\label{chisq_ratio}
\end{equation}
where $p$ is the total number of free parameters in the model and $N$ the total number of datapoints in the fitted time series. Equation~(\ref{chisq_ratio}) allows us to estimate the confidence interval of the parameter $Q$ that depends on the maximum value of the ratio $\chi^{2}/\chi^{2}_{\rm min}$ corresponding to a given confidence limit \citep{Lamptonetal76}. }

The best fit to the ARC2 light curve shows some oscillations of the total $\chi^{2}$ as a function of $Q$, probably related to some residual suppression of the facular modulation by the ARC2 pipeline. Using the original PDC light curve, the oscillations dominate the plot and prevent the determination of the optimal $Q$ value because there is no clearly determined global minimum. This happens because solar-like faculae produce a photometric signal only when they are close to the limb, that is only in limited intervals of the rotational modulation produced by active regions \citep{Lanza16}. Such a tiny signal is easily suppressed by the PDC de-trending leading to insufficient information to constrain the value of $Q$ in the PDC light curve.  On the other hand, the ARC2 pipeline, although still based on the use of the CBVs derived by the PDC pipeline, applies only the first one or a linear combination of the first two CBVs in an attempt to preserve the intrinsic stellar variability on all the accessible timescales. In such a way, most of the facular signal is preserved allowing us to constrain the value of $Q$. The light curve de-trended with the method by \citet{BonomoLanza12}, that removes only long-term trends  by a parabolic fit, gives an even cleaner result. From both the light curves, we derive an optimal value of $Q = 2.4$  that shall be adopted for our analysis. It is different from the value $Q=1.6$ found by \citet{BonomoLanza12} as a consequence of the smaller value of the spot contrast $c_{\rm s}$ used in the present modelling as we verified by running another minimization with the previous $c_{\rm s} = 0.667$.  {Given the impact of the value of $Q$ on our results, in Sect.~\ref{results} and Appendix~\ref{app1}, we shall consider also spot models with $Q=1.0$ and $Q=4.0$, that are well beyond the 95 percent confidence regions in Figs.~\ref{fig2} and~\ref{fig3}, to explore the effect of varying $Q$ on our models. }

The duration of the individual segments of the light curves fitted with the three-spot model has been kept at $\Delta t_{\rm f} = 8.733$ days in all the above models, that is the same value given in \citet{BonomoLanza12}. When we increase $\Delta t_{\rm f}$ by only 2.5 percent, the total $\chi^{2}$ found by minimizing with respect to $Q$ becomes significantly worse {increasing beyond the value corresponding to the 95~percent confidence region as computed with the statistics in Eq.~(\ref{chisq_ratio}).} This suggests that the previous $\Delta t_{\rm f}$ value is still the optimal one. Together with the choice of the same rotation period $P_{\rm rot}$, this has the advantage of yielding spot maps directly comparable with those obtained by \citet{BonomoLanza12}.  {The remarkable sensitivity of the $\chi^{2}$ to $\Delta t_{\rm f }$ is caused by the best fits of some individual time intervals that become significantly worse when adopting a longer $\Delta t_{\rm f}$, likely as a consequence not only of the short spot lifetimes, but also of the difficulty by the numerical optimization routine to reach the same minimum $\chi^{2}$ when the number of datapoints is increased. }

\section{Results}
\label{results}
\subsection{Light curve models}
\label{light_curve_models}
The best fit of the ARC2 light curve without regularization ($\lambda = 0$) was computed with the parameters in Table~\ref{model_param}. The minimum and maximum of the residuals are  $-0.00110$ and $0.00160$, respectively, with an arithmetic mean of $ 1.751 \times 10^{-7}$ in relative flux units. 
The best fit to the distribution of the residuals with a Gaussian has a mean of $-3.428 \times 10^{-6}$ and a standard deviation $\sigma_{\rm ARC2\, 0} = 2.333 \times 10^{-4}$. This value is close to the photometric accuracy of the datapoints indicating that the spot model is generally able to fit the light curve down to the level of the photon shot noise (cf. Sect.~\ref{observations}).

\begin{figure*}
 \centering{
  \includegraphics[width=16cm,height=22cm,angle=0]{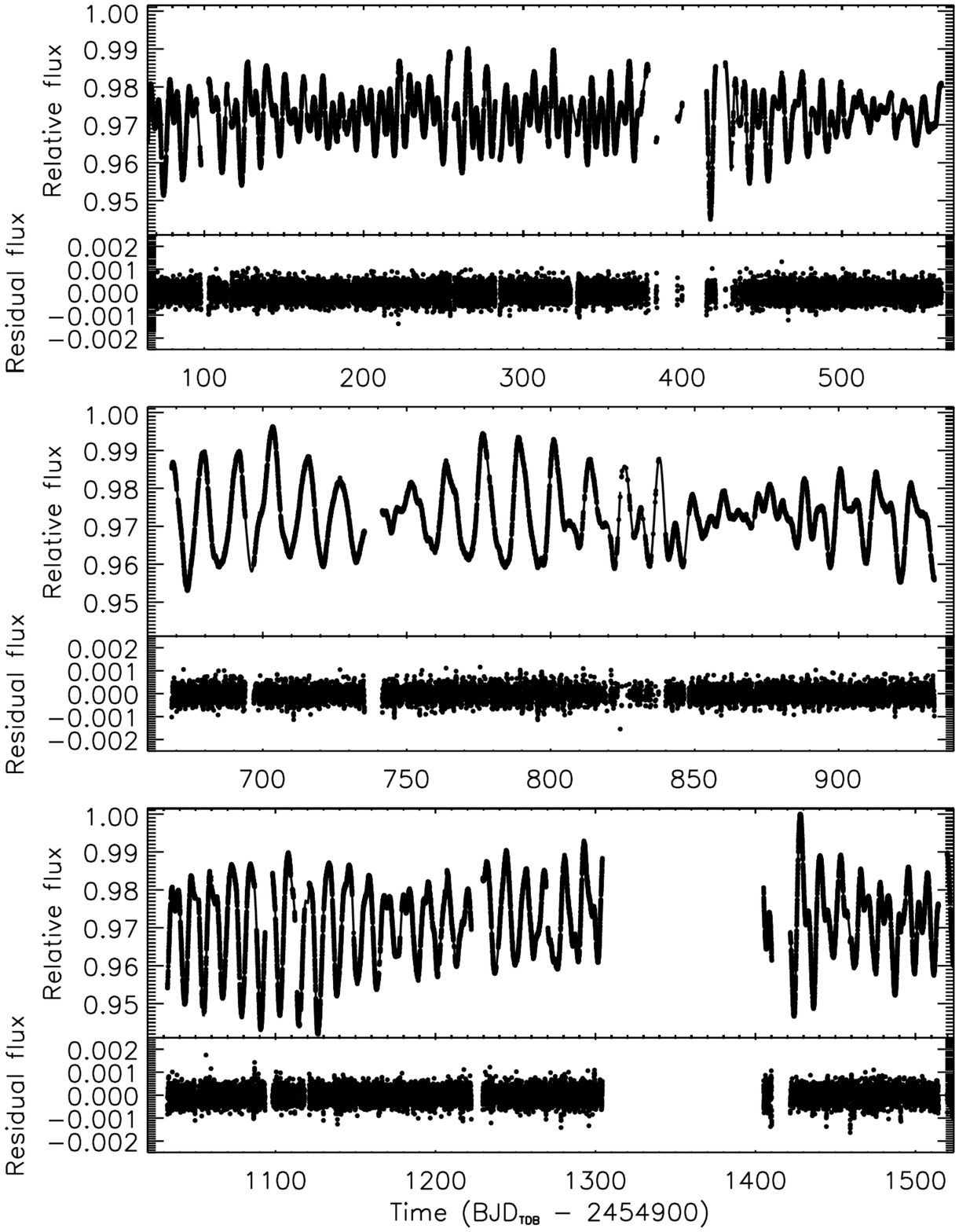}} 
\vspace*{-1.4cm}
   \caption{Top panel: Light curve detrended with the ARC2 pipeline and fitted with our composite maximum-entropy regularized spot model and the parameters listed in Table~\ref{model_param}. The observed flux, normalized to its maximum value, is plotted versus the time (filled dots) and the best fit is superposed (solid line).  Lower panels: the residual of the regularized best fit vs. the time (filled dots).}
              \label{fig6}%
\end{figure*}

\begin{figure}
\hspace*{-1.5cm}
 \centering{
 \includegraphics[width=8cm,height=11cm,angle=90]{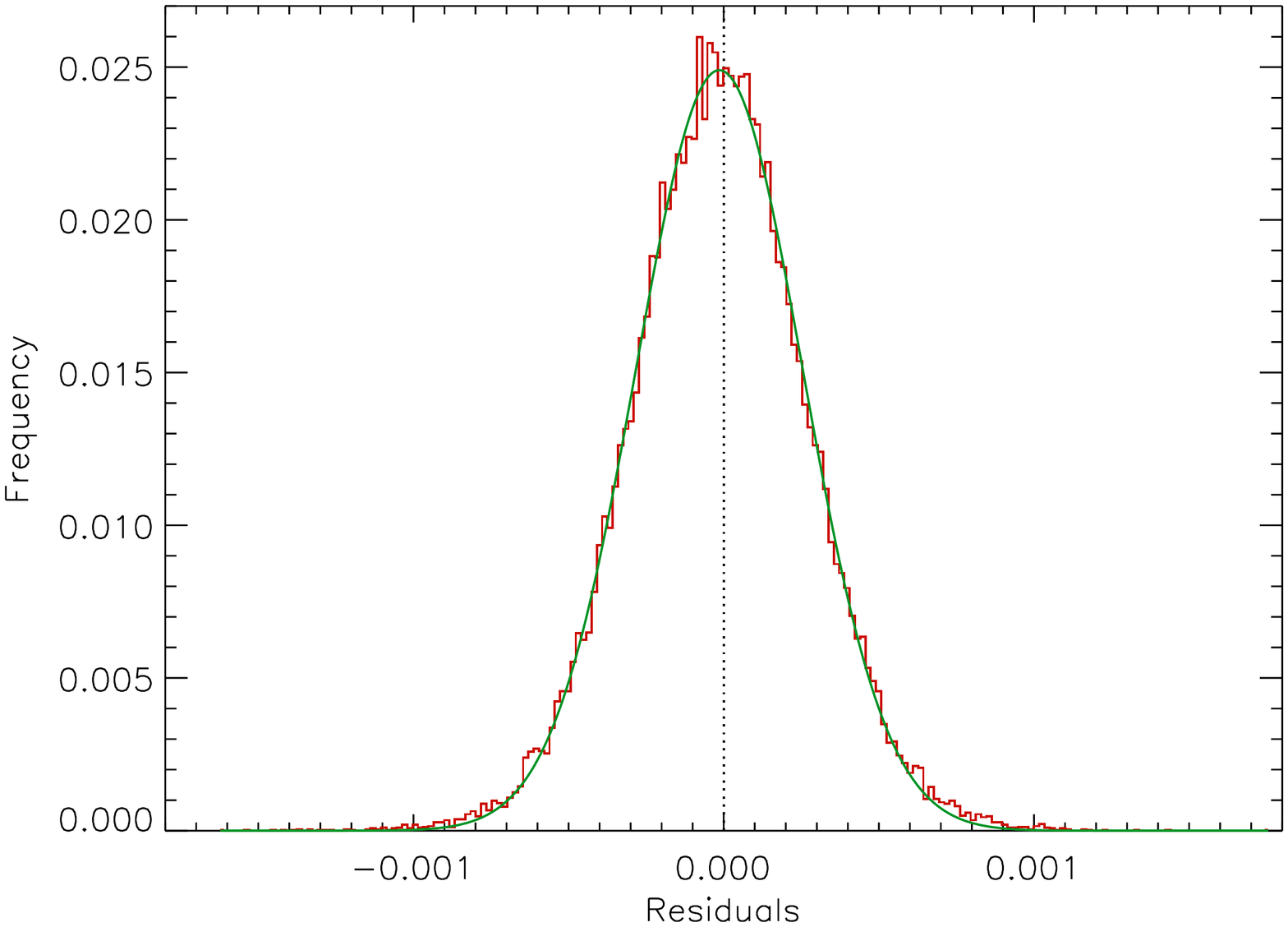}} 
   \caption{Histogram of the distribution of the residuals  of the regularized best fit to the light curve in Fig.~\ref{fig6} (solid red histogram) and its best fit with a Gaussian (solid green line). The vertical dotted line marks the zero value. }
              \label{fig7}%
\end{figure}

The composite regularized best fit obtained with the ARC2 light curve is plotted in Fig.~\ref{fig6}. The light curve and the best fit have been normalized to the maximum of the light curve. The  intervals plotted in the three panels of Fig.~\ref{fig6} have different durations because long gaps have been excluded from our plots with the exception of that in the bottom plot that was left to avoid a fourth plot with too short an interval.  The minimum and maximum residuals are $-0.00163$ and $0.00175$, respectively, with a mean of $-1.325 \times 10^{-5}$ in relative flux units. The distribution of the residuals is plotted in Fig.~\ref{fig7}. The best fit with a Gaussian has a mean of $-1.412\times 10^{-5}$ and a standard deviation of $2.689 \times 10^{-4}$.  The negative mean of the residuals is a consequence of the regularization that tends to reduce the spotted area as much as possible leading to a best fit systematically higher in flux than the data points. The convergence criterion  $|\mu_{\rm reg}| = \sigma_{\rm ARC2\,0}/\sqrt{N}$ is verified within 7 percent in all the cases with more than 75 percent of the individual intervals verifying it within 2 percent. 

The best fit of the light curve obtained with the de-trending approach of \citet{BonomoLanza12} and without regularization ($\lambda =0$) was computed with the parameters listed in Table~\ref{model_param}. The residuals of the best fit range between $-0.00230$ and 0.00241  with a mean of $3.491 \times 10^{-7}$ in relative flux units. The largest residuals are found close to data gaps and are probably associated with the residual systematics  before and after the gaps that the approach by \citet{BonomoLanza12} is not capable of correcting as efficiently as the ARC2 pipeline.  The best fit with a Gaussian to the distribution of the residuals has a mean of $-2.631 \times 10^{-6}$ and a standard deviation $\sigma_{\rm BL\, 0}=2.504 \times 10^{-4}$ in relative flux units.  

\begin{figure*}
 \centering{
  \includegraphics[width=16cm,height=22cm,angle=0]{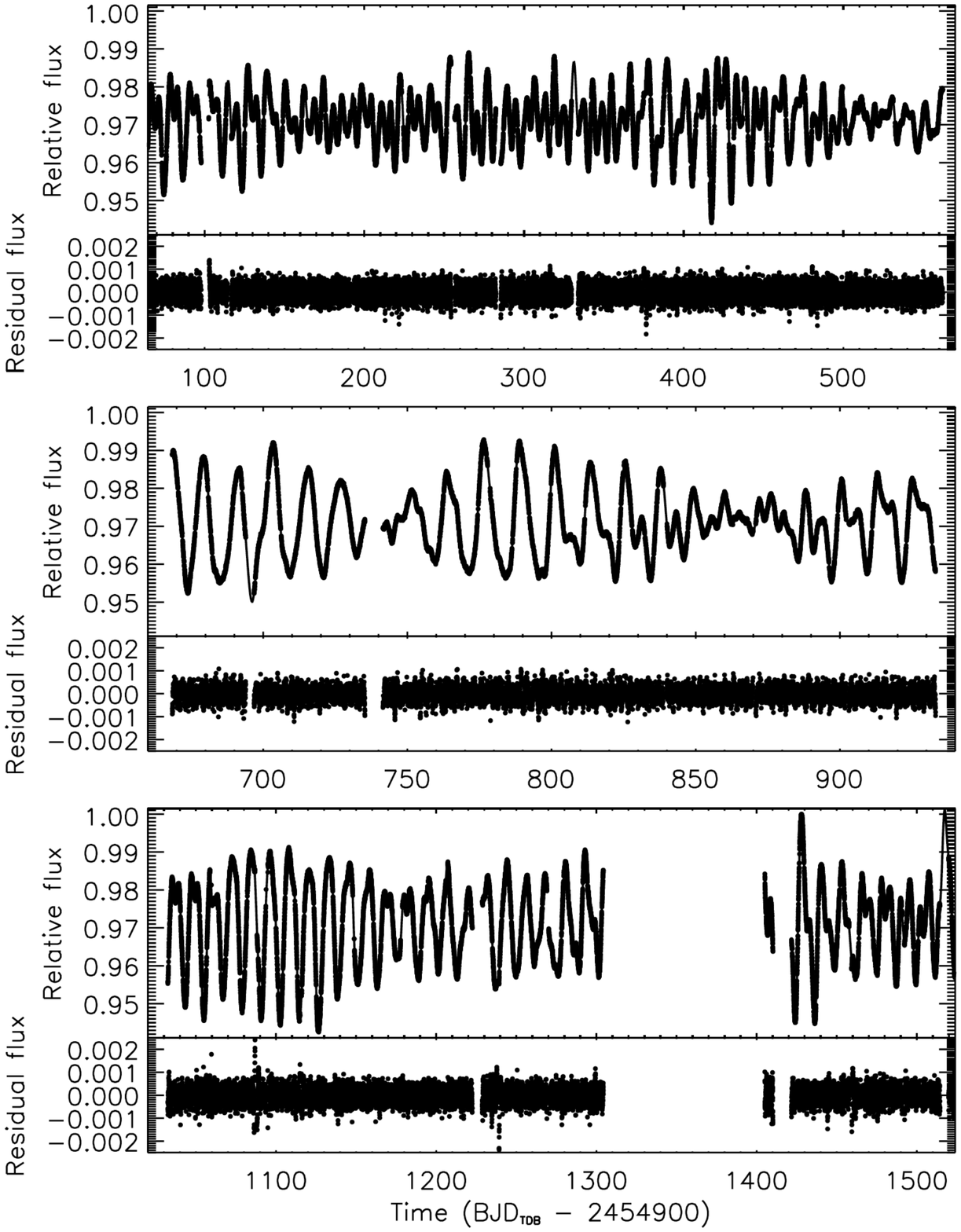}} 
 \vspace*{-1.4cm}
\caption{Same as Fig.~\ref{fig6} for the light curve de-trended with the approach of \citet{BonomoLanza12}. }
              \label{fig4}%
\end{figure*}

\begin{figure}
\hspace*{-1.5cm}
 \centering{
 \includegraphics[width=8cm,height=11cm,angle=90]{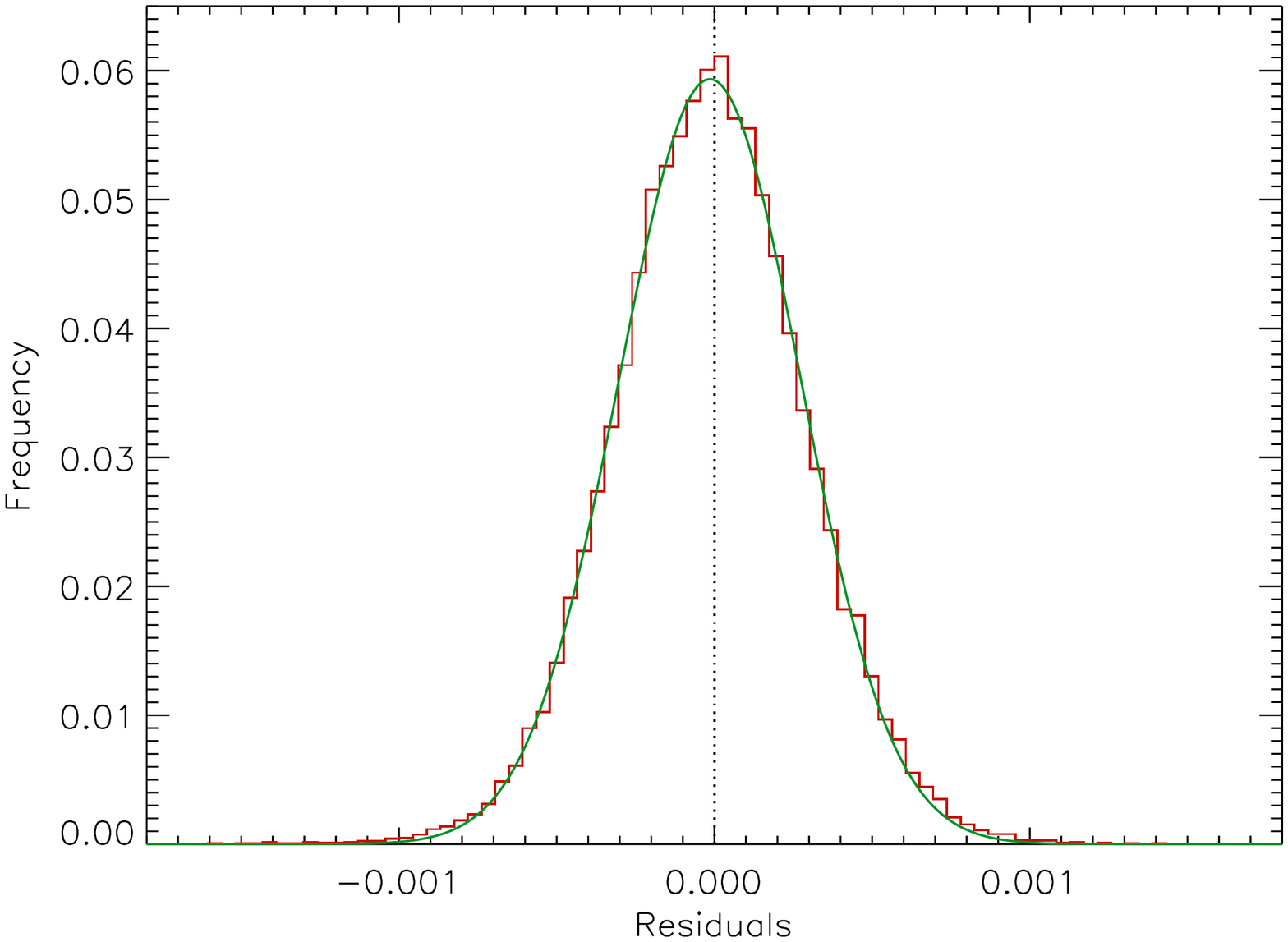}} 
 \caption{Same as Fig.~\ref{fig7} for the light curve de-trended as in \citet{BonomoLanza12} and plotted in Fig.~\ref{fig4}. }
              \label{fig5}%
\end{figure}

The composite regularized best fit of the light curve obtained with the approach of \citet{BonomoLanza12} is plotted in Fig.~\ref{fig4}. The light curve and the best fit have been normalized to the maximum flux of the light curve.  The minimum and the maximum of the residuals are $-0.00603$ e $ 0.00260$, respectively, while the  mean of the residuals is $-1.326 \times 10^{-5}$ in relative flux units.    The distribution of the residuals is plotted in Fig.~\ref{fig5} together with a Gaussian best fit having a mean of $-1.415\times 10^{-5}$ and a standard deviation of $3.067 \times 10^{-4}$.  While the mean is comparable with that of the fit to the ARC2 light curve, the standard deviation is about 10 percent larger as indicated by the greater fraction of relatively larger residuals. The convergence criterion $|\mu_{\rm reg}| = \sigma_{\rm BL\,0}/\sqrt{N}$ is verified with a maximum deviation of 6 percent for all the individually fitted intervals of duration $\Delta t_{\rm f} = 8.733$ days, with more than 75 percent of the intervals verifying the criterion within 3 percent. 

In contrast with the distribution of the residuals of the light curve fitted by \citet{BonomoLanza12}, that showed a positive tail in excess of the Gaussian best fit, the present distribution is well fitted by a Gaussian and has a symmetric shape. This difference is likely due to the improved correction of the residual systematics in the latest release of Kepler data. 

\begin{figure}
\hspace*{-1.5cm}
 \centering{
 \includegraphics[width=8cm,height=11cm,angle=90]{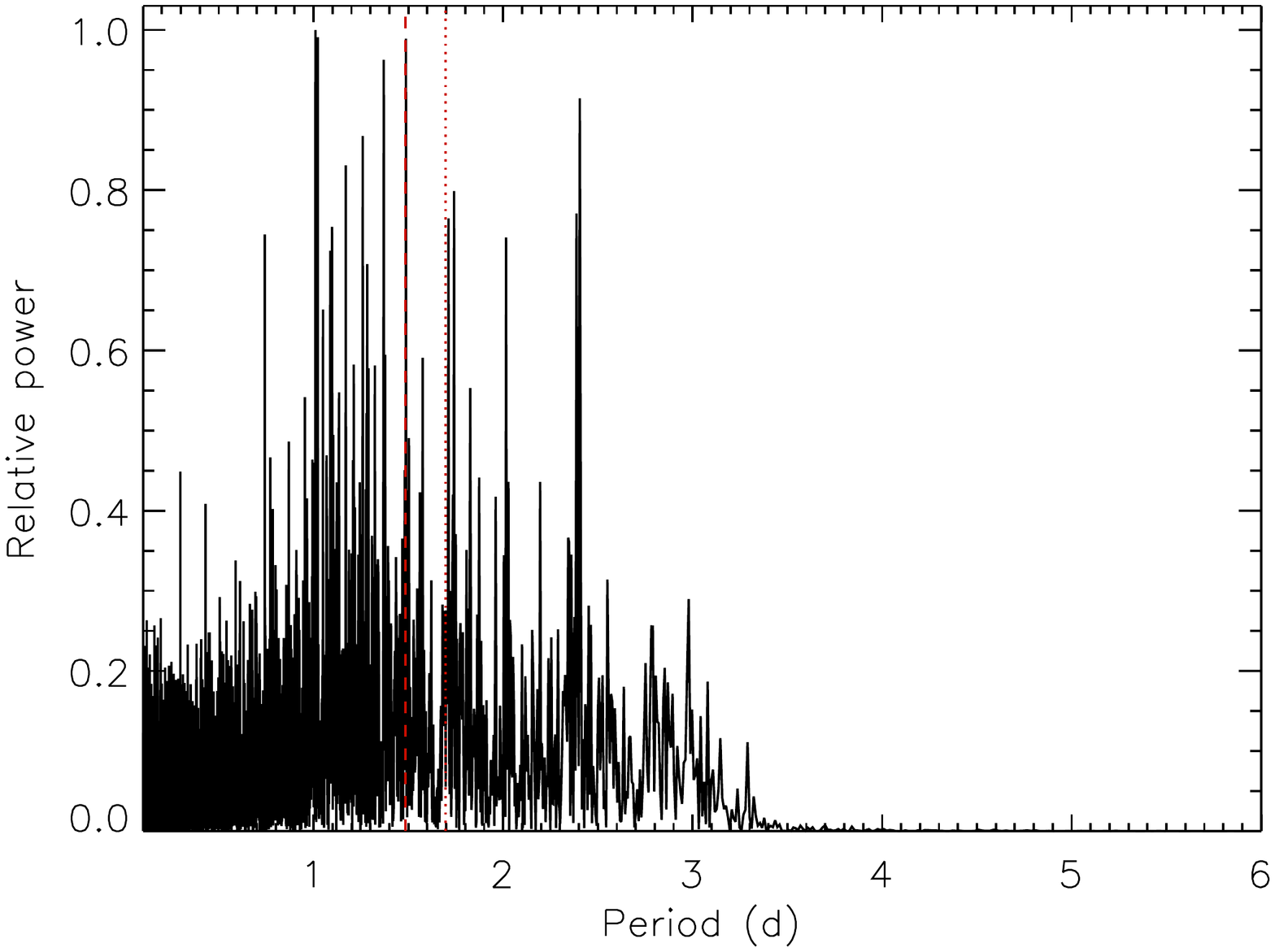}} 
   \caption{Generalized Lomb-Scargle periodogram of the residuals of the unregularized best fit to the light curve de-trended with the approach of \citet{BonomoLanza12}. The vertical red dashed line indicates the orbital period of Kepler-17b, while the red dotted line indicates the synodic period (see the text).  }
              \label{fig7gls}%
\end{figure}

It is interesting to explore possible periodicities in the time series of the residuals of the light curve best fits. To this purpose, we analyze the residuals of the unregularized best fit because they have no systematic bias introduced by the regularization. In Fig.~\ref{fig7gls}, we plot  the Generalized Lomb-Scargle periodogram \citep[hereafter GLS, see][]{ZechmeisterKuerster09} of the residuals to the light curve de-trended with the approach of \citet{BonomoLanza12}, chosen because it has more datapoints and less gaps than the ARC2 light curve. {Given that the removal of the transits may affect the power spectrum, we checked the spectral window of the time series finding low sidelobes at frequencies of $\sim 2$ and $\sim 3.25$ day$^{-1}$ that are away from the orbital frequency of 0.672~day$^{-1}$, showing that the transit removal does not significantly affect our periodogram. Actually, the relative duration of the discarded intervals is 12.7 percent of each orbital period that explains why the impact on the spectral window is limited when applying the GLS periodogram, specifically designed to treat time series with gaps.} 

We find that our spot modelling accounts for the light modulations with timescales longer than $\sim 3.5$ days as indicated by the almost complete disappearance of power in the periodogram for periods longer than that limit.  There is a peak at the orbital period with a false-alarm probability of $\sim 10^{-8}$ as estimated with the analytic formula of \citet{ZechmeisterKuerster09}. This suggests a  variability with the orbital phase that can be attributed to  light reflection and secondary eclipses  as discussed by \citet{Bonomoetal12}. However, we are not in the position to characterize such effects because we are analysing only long-cadence data the time sampling of which is too coarse to give precise information on these phenomena, in particular on the secondary eclipses. We also indicate in Fig.~\ref{fig7gls} the synodic period $P_{\rm syn}$ calculated as $P_{\rm syn}^{-1} = | P_{\rm orb}^{-1} - P_{\rm rot}^{-1}|$, where we adopted a mean rotation period $P_{\rm rot} = 12.01$~days. Variability with the synodic period could be an indication of magnetic star-planet interactions \citep{Lanza08,Lanza12}. In the present case, the peaks close to $P_{\rm syn}$ are not particularly prominent suggesting that there is no detectable effect of the planet on stellar variability in the present dataset. Similar results are obtained from the analysis of the residuals of the regularized best fits, but they are not shown here. 

\subsection{Longitude distribution of active regions, comparison with occulted spots, and differential rotation}
\label{spot_active_longitudes}

In Fig.~\ref{fig9} we plot the distribution of the filling factor of the starspots $f$ vs. the longitude and the time for the regularized spot maps obtained with the ARC2 light curve. The origin of the longitude is at the meridian pointing toward the Earth on BJD$_{\rm TDB}$ 2454964.512 and the longitude is increasing in the same direction as the stellar rotation and the orbital motion of the planet. The reference frame is rotating with the star with a period $P_{\rm rot} = 12.01$~days as in \citet{BonomoLanza12}.  In the following, we measure the time $t$ starting from a reference epoch, that is, we introduce a time $t^{\prime} = t - 2454900.0$ measured in BJD$_{\rm TDB}$.  

Spots occulted by the planet during transits have been modelled by \citet{Valioetal17} by analysing  the short-cadence Kepler photometry whose time series began after the first transit detection, that is, approximately after $t^{\prime} \sim 290$ days.  {The belt covered by the planet during its transits extends from $-7.0^{\circ}$ to $8.4^{\circ}$ in latitude, assuming that the transit chord covers the northern hemisphere of the star, given that the impact parameter  is $b = 0.012 \pm 0.010$ and the radius of the planet $0.1335 \pm 0.0001$ stellar radii according to \citet{Maxted18}.  A similar result is obtained with the transit model by \citet{Valioetal17} that took into account the systematic effects of the spots on the transit profile obtaining a slightly larger radius than Maxted's, that is, $0.138 \pm 0.001$ and an impact parameter $b = 0.10 \pm 0.01$ stellar radii, leading to an occulted belt between $-3.6^{\circ}$ and $13.8^{\circ}$ in latitude, again assuming a transit chord on the northern hemisphere.} In Fig.~\ref{fig9}, the  occulted spots  are marked as white open circles the size of which is proportional to their flux deficit defined as $D=\pi r^{2}_{\rm s} (1-I_{\rm spot}/I)$, where $r_{\rm s}$ is the radius of the spot as derived by the duration of its occultation and $I_{\rm spot}/I$ the ratio of its specific intensity to the unperturbed intensity as derived from the height of the photometric anomaly produced by the spot itself (the "bump" along the residual transit profile). 

The map of the distribution of the spot filling factor as derived from the  light curve de-trended with the approach of \citet{BonomoLanza12} is displayed in Fig.~\ref{fig8} together with the spots occulted by the planet during its transits. The ARC2 pipeline discarded the datapoints of an interval centred around $t^{\prime} \sim 400$~days that were instead retained by the simpler de-trending algorithm applied by \citet{BonomoLanza12}, therefore the corresponding map shows only three major gaps.

The two maps in Figs.~\ref{fig9} and~\ref{fig8} are remarkably similar.  This is due to the fact that the maps are based on best fits to short individual time intervals of $\Delta t_{\rm f} = 8.733$~days, along which the light variations are comparable in the two time series, thus the different trends of the light curves on longer timescales do not strongly affect the spot distributions. However, we see that the map based on the ARC2 light curve shows a slight preference for a greater filling factor, in particular for values closer to the maximum, as indicated by the more extended red and orange areas in Fig.~\ref{fig9}. {In Appendix~\ref{app2} and Fig.~\ref{fig8_detail}, we show an enlargement of Fig.~\ref{fig8} to better show the migration of the spot pattern and the association between spots mapped from the out-of-transit light curve and spots occulted during transits.

Considering Figs.~\ref{fig9} and~\ref{fig8}, during the first time interval ($ t^{\prime} \la 550$), two main active longitudes are apparent, one beginning at $\approx 50^{\circ}$ longitude and slowly migrating to $\approx 0^{\circ}$, the other beginning at $\approx 200^{\circ}$ and staying approximately stationary in the adopted reference frame. Another active longitude appears in between, characterized by a remarkable intermittency.  Individual starspots have a duration ranging from $\approx 10$ to $\approx 40- 50$ days, that is remarkably shorter than the duration of the active longitudes (cf. Fig.~\ref{fig8_detail}). 
This pattern is closely similar to that mapped by \citet{BonomoLanza12} using the Kepler long-cadence data available at that time (cf. their Fig.~4).  

{To measure the association between the longitude distribution of the spots as derived from the out-of-transit light curves and the distribution as mapped from the occultations during transits, we define the cross-correlation coefficient $\rho_{\rm cc}$ as:
\begin{equation}
\rho_{\rm cc} (\ell) = \left\{ \begin{array}{ll} 
 \frac{\sum_{k=1}^{N_{\rm L}-| \ell |} (s_{\rm oot}(k+|\ell |) - \overline{s}_{\rm oot})(s_{\rm occ}(k)-\overline{s}_{\rm occ})}{\sqrt{\sum_{k=1}^{N_{\rm L}} (s_{\rm oot}(k) - \overline{s}_{\rm oot})^{2} (s_{\rm occ}(k)-\overline{s}_{\rm occ})^{2}}} & \mbox{for $\ell < 0$} \\
 \frac{\sum_{k=1}^{N_{\rm L}-\ell} (s_{\rm oot}(k) - \overline{s}_{\rm oot})(s_{\rm occ}(k+\ell)-\overline{s}_{\rm occ})}{\sqrt{\sum_{k=1}^{N_{\rm L}} (s_{\rm oot}(k) - \overline{s}_{\rm oot})^{2} (s_{\rm occ}(k)-\overline{s}_{\rm occ})^{2}}} & \mbox{for $\ell \geq 0$}, 
 \end{array}
 \right.
 \label{cc_eq}
\end{equation}
where $s_{\rm oot}$ is the out-of-transit longitude distribution of the spot filling factor $f$ and $s_{\rm occ}$ the distribution of $D$ of the occulted spots, both mapped onto $N_{\rm L}=20$ equal longitude bins of $18^{\circ}$;  the overbar indicates the mean value; and $\ell \in [-10, 10]$ is the lag index, the longitude lag being given by $\Delta \lambda = 18^{\circ} \times \ell$. The distributions $s_{\rm occ}$ and $s_{\rm oot}$ are treated as circular datasets, that is, they repeat themselves beyond an interval of $360^{\circ}$. Because $\Delta t_{\rm f} < P_{\rm rot}$, we consider the mean of two consecutive out-of-transit spot distributions and smooth them to a resolution of $54^{\circ}$ to derive $s_{\rm oot}$. Similarly, to have a complete longitude coverage along the chords occulted by the planet, we average the distributions of the occulted spots along four consecutive transits to compute $s_{\rm occ}$. The association between the two distributions is measured by the cross-correlation coefficient at zero lag, that is, $\rho_{\rm cc} (0)$ that is plotted vs. the time in Fig.~\ref{correlation_in_out_transits} for our two light curves. The correlation is zero for $t^{\prime} \la 290$ because no transit were observed. Note that $\rho_{\rm cc}(0)$ is equal to the Pearson linear correlation coefficient $r$ between $s_{\rm oot}$ and $s_{\rm occ}$ as defined in Sect.~14.5 of \citet{Pressetal02}. Estimating the significance of the correlation, that is, the probability of obtaining the given $r$ or a larger one in the case of a chance association,  is difficult because the statistical distributions of the correlated variables are in general not known. A good alternative is to resort to the Spearman or rank-correlation coefficient $r_{\rm s}$ for which an analytic evaluation is possible \citep[cf.][ Sect.~14.6]{Pressetal02}. Therefore,  we can use $r_{\rm s}$ to compute the significance of $\rho_{\rm cc}(0)$.

We can apply Eq.~(\ref{cc_eq}) also to evaluate the migration of the spot pattern between two consecutive out-of-transit spot longitude distributions. Specifically, the longitude lag $\Delta \lambda$ that maximizes the cross-correlation $\rho_{\rm cc}$ between consecutive out-of-transit spot distributions  can be used to quantify the longitude migration of the spot pattern occurred between them. This migration is assumed to be produced by the differential rotation when the most prominent spots are not rotating with the period of $12.01$~days assumed for the reference frame. In Fig.~\ref{lag_migration}, the migration rate obtained from the derivative of $\Delta \lambda$  as a function of the time is plotted vs. the time itself.  

Considering Figs.~\ref{fig9} and~\ref{fig8}, we see that {several} spots occulted during transits are concentrated around the active longitudes found by the out-of-transit spot modelling with the coincidence being better for the two more persistent longitudes {and during some time intervals, in particular when $\rho_{\rm cc}(0) \ga 0.25$. } This coincidence provides an independent confirmation of our spot modelling approach and indicates that {some of the} occulted spots and the active longitudes mapped by the out-of-transit light curve are at a similar latitude. The migration of the trails of the occulted spots towards positive longitudes indicates that they are rotating slighly faster than the main active longitude with a period of $\sim 11.90\pm 0.04$~days. Note that the longitude resolution of the maps of the occulted spots can be as small as a few degrees, while that of the spots mapped from the out-of-transit light curve reaches only $\approx 50^{\circ}$ in the best cases, thus accounting for the lack of a complete coincidence between the two maps \citep[cf.][]{Lanzaetal07,SilvaValioLanza11} and the relatively low significance of the correlation that ranges from $\sim 0.1$ to $\sim 0.4$ for $\rho_{\rm cc}(0) \sim 0.25$ as derived from the analytical method introduced above.  

The best correspondence between the out-of-transit and the occulted spot distributions is found for $t^{\prime} \approx 300, 900$, and $1100$ as indicated in Fig.~\ref{correlation_in_out_transits} by $\rho_{\rm cc}(0) \ga 0.35$ that corresponds to an analytical significance better than 0.1. At those times, we see active longitudes in Figs.~\ref{fig9} and~\ref{fig8}, while the migration rate of the out-of-transit spot distributions is close to zero or fluctuates between zero and $+1.5$ deg/day in Fig.~\ref{lag_migration}, indicating that the out-of-transit light curve is dominated by low latitude spots, mostly occulted during transits. On  the other hand, when $\rho_{\rm cc}(0) \la 0.2$, active longitudes are not apparent in the adopted  reference frame and we see mostly negative migration rates as, for example, for $ 750 \la t^{\prime} \la 850$  or $ 1400 \la t^{\prime} \la 1500$ (cf. Fig.~\ref{lag_migration} and Fig.~\ref{fig8_detail}). The minimum migration rate is of $-$(3-4)~deg/day, if we consider  couples of consecutive similar measurements and exclude more extreme, isolated values which have smaller correlation coefficients as indicated by the smaller sizes of their plotted symbols. This minimum migration rate corresponds to a rotation period of $13.35-13.85$ days that gives a relative differential rotation of $\Delta P_{\rm rot}/P_{\rm rot} \simeq 0.14 \pm 0.05$ considering an error of $\pm 20^{\circ}$ in the measurement of the lag between successive distributions and basing the determination on  two consecutive measurements. The differential rotation is solar-like, that is, the lower latitudes rotate faster than the higher latitudes as we deduce from the occultation of the faster rotating spots during transits.

The  regular increase of the migration rate as observed in Fig.~\ref{lag_migration} during $750 \la t^{\prime} \la 850$ or $1400 \la t^{\prime} \la 1500$ can be interpreted as a consequence of the evolution of the high-latitude spots that produce the negative values of the rate itself. When those spots first appear, they dominate the correlation between successive spot distributions producing the largest negative deviations from the angular velocity corresponding to the rotation period of the reference frame of 12.01~days. Their subsequent decay makes the spots at lower latitudes increasingly more relevant in the cross-correlation until the migration rate crosses the zero value and becomes positive when the latter completely dominate the cross-correlations. From the duration of such phases of $\approx 100$~days, we deduce that high-latitude spots have a maximum lifetime of about three months. Alternatively, individual high-latitude spots may have a shorter lifetime (cf. Fig.~\ref{fig8_detail}), while it is the latitude of the activity belt that steadily migrates towards the equator as we observe in the solar cycle. In this case, we can trace three cycles in Fig.~\ref{lag_migration} separated by $\approx 400$ and $\approx 600$~days, respectively. These periods agree with the period in the total area of the occulted spots as found by \citet{EstrelaValio16} who reported a modulation with a period of $490 \pm 100$~days.

In Fig.~\ref{lag_migration}, we see some very fast changes in the migration rate of the starspots. This is similar to the observations of the solar rotation period as derived from disc-integrated tracers such as the chromospheric Ca~II~H\&K lines. In those timeseries, the onset of a new cycle is marked by an abrupt increase of the period as activity disappears at low latitudes and re-appears at higher latitudes \citep[e.g.][]{DonahueKeil95,HempelmannDonahue97}.

We explore the effect of varying the parameters of our spot modelling in Appendix~\ref{app1} finding that a decrease of the contrast of the spots $c_{\rm s}=I_{\rm spot}/I$ or of the facular-to-spotted area ratio $Q$ have the highest impact on our measurement of the differential rotation reducing its amplitude by approximately a factor of two for the extreme values of those parameters. 

\begin{figure*}
 \centering{
 \includegraphics[width=17cm,height=21cm,angle=0]{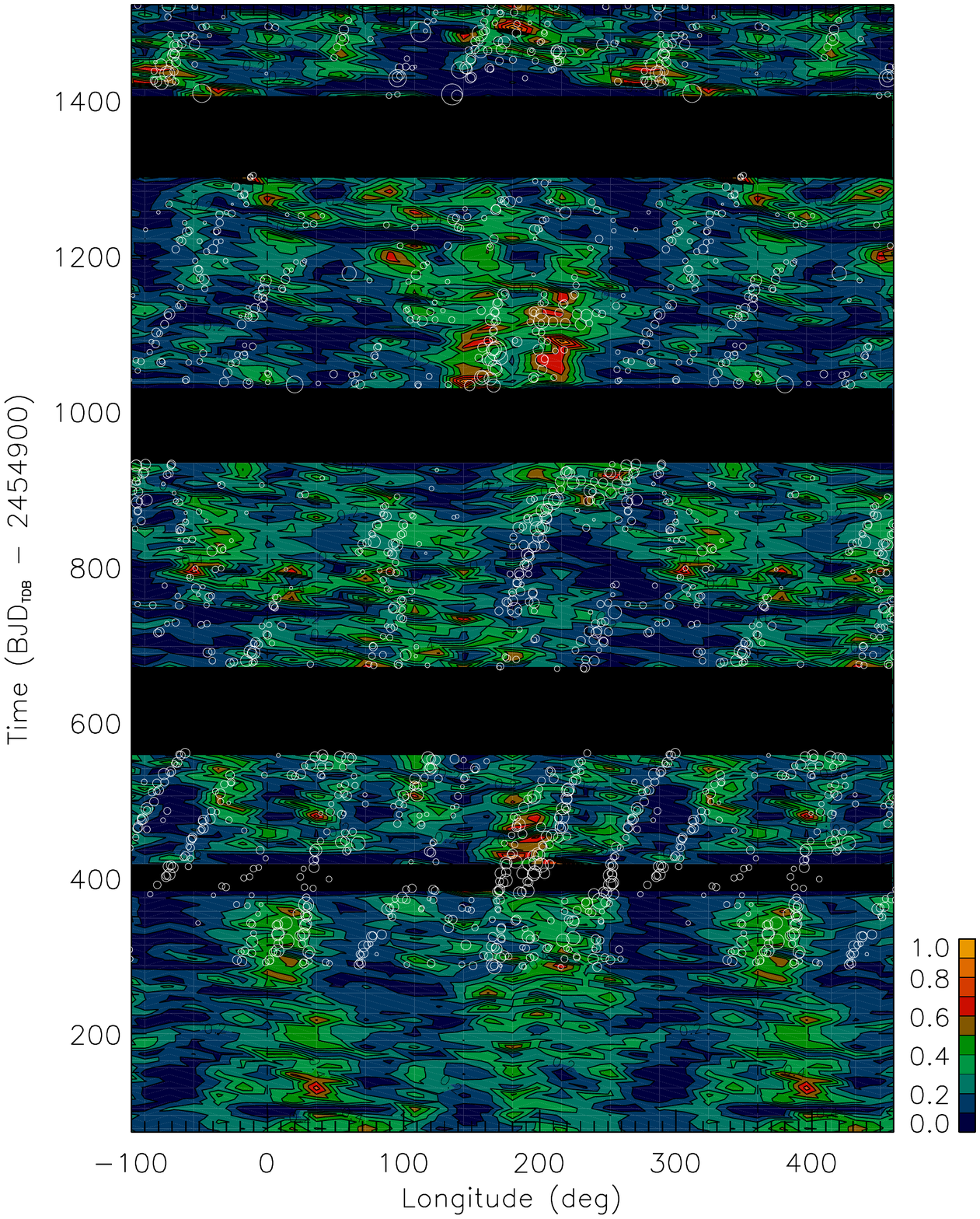}} 
 \vspace*{-1.7cm}
  \caption{Distribution of the spot filling factor vs. the longitude and time as derived by our maximum-entropy spot model of the long-cadence ARC2 light curve. The maximum of the filling factor is indicated by the yellow-orange colour, while the minimum by dark blue (see colour scale in the lower right corner). Note that the longitude scale is repeated beyond the $[0^{\circ}, 360^{\circ}]$ interval to better follow the migration of the spot features. White circles mark the longitude and  time of the spots occulted by the planet during transits as detected by modelling transit profiles observed in short cadence \citep{Valioetal17}. Their size is proportional to their flux deficit $D$ as defined in Sect.~\ref{spot_active_longitudes} (see the text). Data gaps without enough observations to compute a spot map along the individual $\Delta t_{\rm f}$ intervals are indicated by a black band. Note that short-cadence transit data are available during the first gap of the light curve extracted by the ARC2 pipeline.}
              \label{fig9}%
\end{figure*}
\begin{figure*}
 \centering{
 \includegraphics[width=17cm,height=21cm,angle=0]{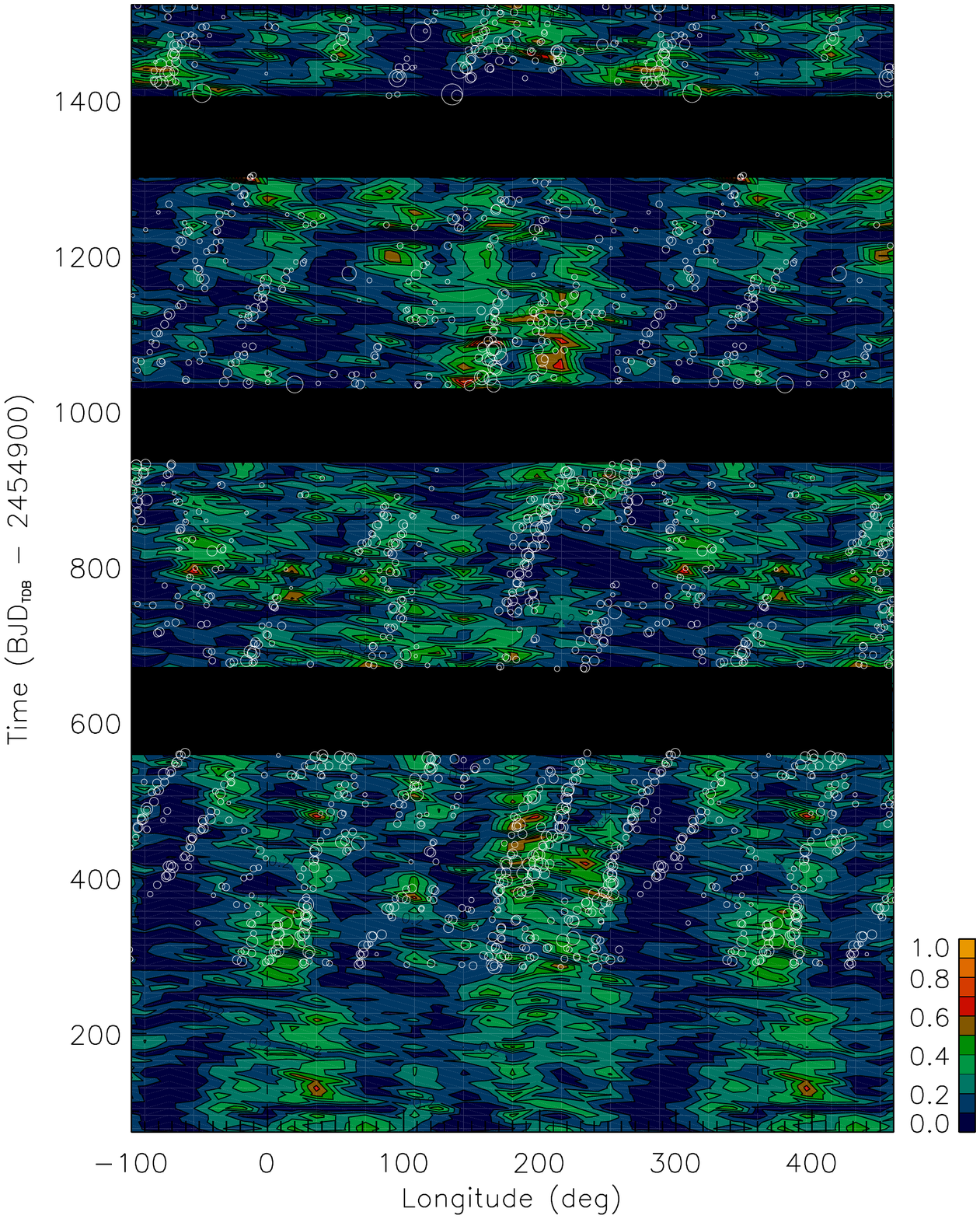}} 
 \vspace*{-1.7cm}
\caption{Same as Fig.~\ref{fig9} for the spot modelling of the  light curve de-trended with the method of \citet{BonomoLanza12}.}
              \label{fig8}%
\end{figure*}
\begin{figure}
\hspace*{-1.cm}
 \centering{
 \includegraphics[width=8cm,height=10cm,angle=90]{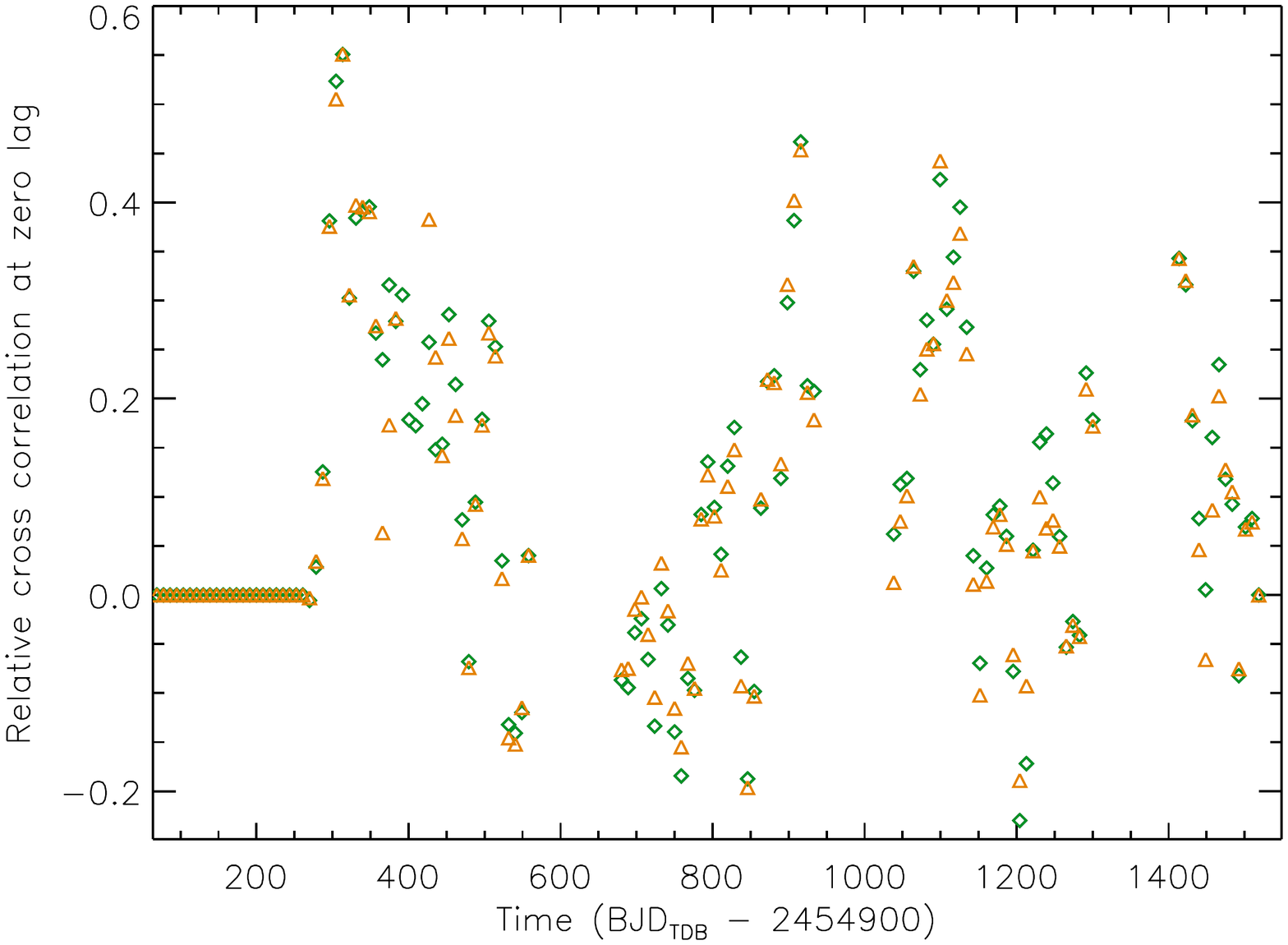}} 
 \vspace*{-1.cm}
  \caption{Cross-correlation coefficient at zero lag (see text) between the distributions of the starspots as obtained from the ME models of the out-of-transit light curves and the spots occulted during transits as mapped by \citet{Valioetal17}. The considered ME spot longitude distributions are those obtained from the light curve of Bonomo \& Lanza (green diamonds) and the ARC2 light curve (orange triangles). }
              \label{correlation_in_out_transits}%
\end{figure}
\begin{figure}
\hspace*{-1.cm}
 \centering{
 \includegraphics[width=8cm,height=10cm,angle=90]{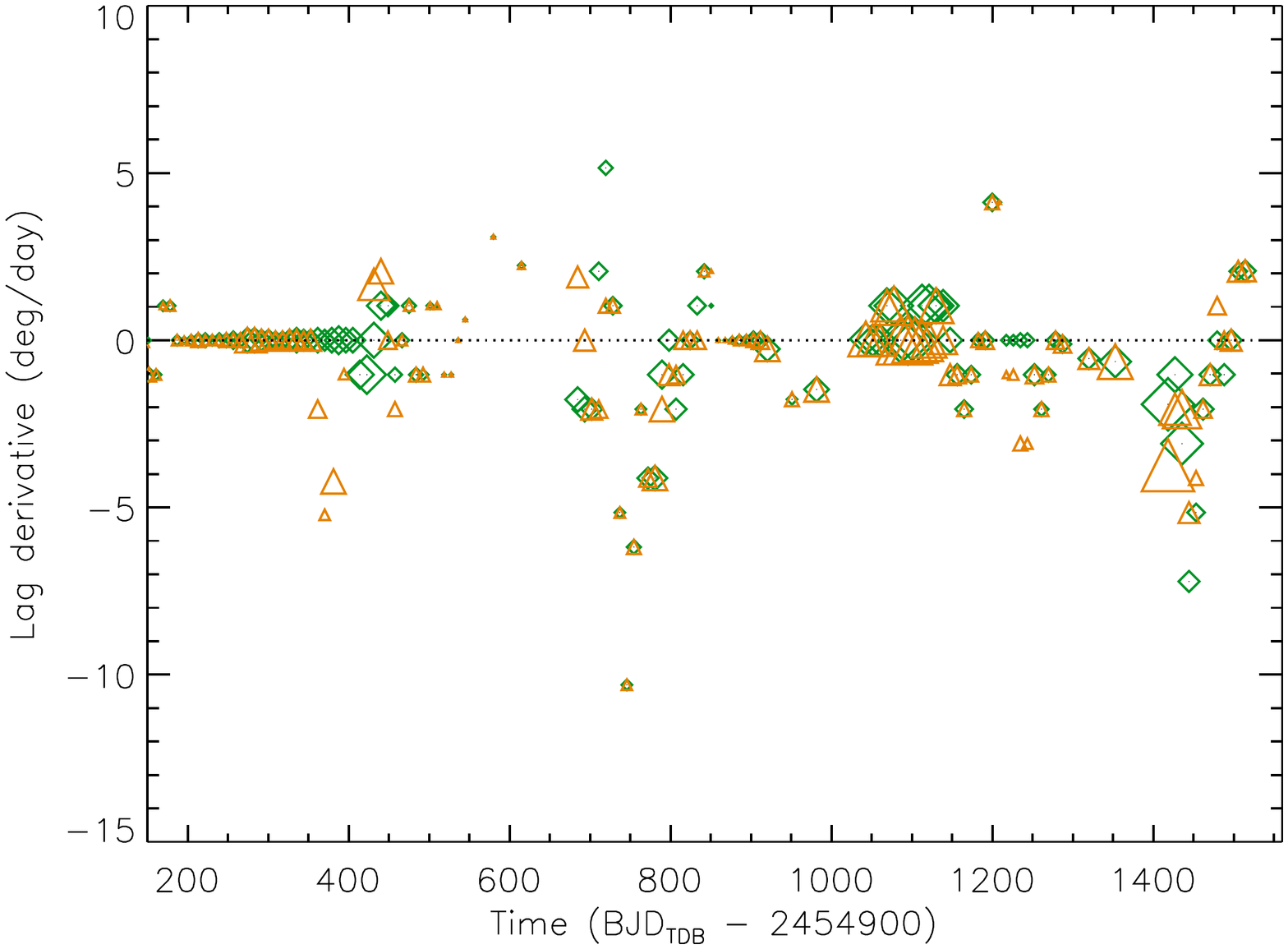}} 
 \vspace*{-1.cm}
  \caption{Migration rate between consecutive spot pattern distributions as derived from  the Bonomo \& Lanza light curve (green diamonds) and the ARC2 light curve (orange triangles). The size of the symbols is proportional to the cross-correlation coefficient $\rho_{\rm cc}$ (cf. Eq.~\ref{cc_eq}). }
              \label{lag_migration}%
\end{figure}
 
\subsection{Variation of the spotted area}
\label{spotted_area_variation}

Our spot modelling of the out-of-transit light curve allows us to determine the variation of the total spotted area vs. the time by integrating the filling factor over the longitude.  The error is estimated from the photometric accuracy of the datapoints. The presence of gaps inside each individually fitted interval of duration $\Delta t_{\rm f}$ affects the total area because the maximum entropy regularization drives the solution towards the minimum spotted area compatible with the data, thus reducing the filling factor at the longitudes that are in view during the gaps in the light curves. 

To reduce the impact of this effect on the variation of the total spotted area, we measured the presence of significant gaps along each interval $\Delta t_{\rm f}$. We divided each interval into five equal subintervals and counted the number of datapoints into each subinterval $n_{i}$, with $i=1,..,5$ numbering the subinterval. A measure $\delta$ of the inhomogeneous distribution of the datapoints along the interval $\Delta t_{\rm f}$ is defined as $\delta \equiv [\max (n_{i})-\min (n_{i})]/ \max (n_{i})$. In the case of the area values obtained from the light curve with the de-trending of \citet{BonomoLanza12}, the intervals with $\delta > 0.2$ are discarded giving a total of 83 area measurements unaffected by the gaps over a total of 135 intervals. {Note that, as the intervals with $\delta > 0.2$ have on the average $\sim 4$~percent less datapoints, they show a similar systematic decrease of the spot coverage values. With a mean spot coverage of $\sim 0.068$, this amounts to a systematic difference of $\sim 2.7 \times 10^{-3}$ that is comparable with the amplitude of the modulation we detect in the spot coverage itself (see below). Therefore, we choose $\delta =0.2$ as our acceptance threshold to avoid systematic errors comparable with the amplitude we intend to measure.  

For the spot coverage obtained by the best fits to the ARC2 light curve, we had to relax the acceptance criterion and discarded only the intervals with $\delta > 0.25$  to have a comparable number of coverage measurements, specifically,  70 acceptable values over a total of 133 time intervals.} Note that the ARC2 light curve has less datapoints and more gaps than the light curve de-trended with the approach of \citet{BonomoLanza12}, thus we base our analysis mainly on the latter light curve. 

The plot of the total spotted area vs. the time for this light curve is shown in Fig.~\ref{fig10} together with the best fitting sinusoid with a period of 47.906 days as obtained by the GLS periodogram. {The area values put in phase with that period are shown in Fig.~\ref{fig10phased}.} The false-alarm probability (FAP) estimated with the analytic formula proposed by \citet{ZechmeisterKuerster09} is 0.0166. {By performing $10\, 000$ shuffling of the area values of the time series, we estimate a FAP of 0.05, not too different from the analytic estimate.} The plot in Fig.~\ref{fig10} shows that the 48-d oscillation is particularly evident for $t^{\prime} < 600$~days, possibly around $t^{\prime} \sim 800$~days, and in the latest part of the time series, that is for $t^{\prime} > 1100-1200$~days, although with a varying amplitude. 

To trace this varying periodicity, we apply a Morlet wavelet with the same parameters as in \citet{BonomoLanza12}. The amplitude of the wavelet vs. the period and time is plotted in Fig.~\ref{fig11}. The relative maxima of the power are concentrated around a period of $\sim 50$~days, although there is sometimes power at periods around 30~days. The gaps in the  time series affect the Morlet wavelet and can account for the secondary maxima at different periods.   A comparison of the wavelet map in Fig.~\ref{fig11} with that in Fig.~7 of \citet{BonomoLanza12} shows the same overall structure, although the new map extends for a longer time interval and suggests a re-appearance of the periodicity at $\sim 50$~days close to the end of the time series.  

{We explore the impact of the variation of our model parameters on the spot coverage in Appendix~\ref{app1} considering different values of the limb-darkening coefficients, spot contrast $c_{\rm s}=I_{\rm spot}/I$, and facular-to-spotted area ratio $Q$. We find that the $\sim 48$-d periodicity is retrieved in all the cases, although its false-alarm probability becomes larger for non-optimal values of the $Q$ parameter or the largest values of the spot contrast $c_{\rm s}$.}
\begin{figure}
\hspace*{-1.5cm}
 \centering{
 \includegraphics[width=8cm,height=11cm,angle=90]{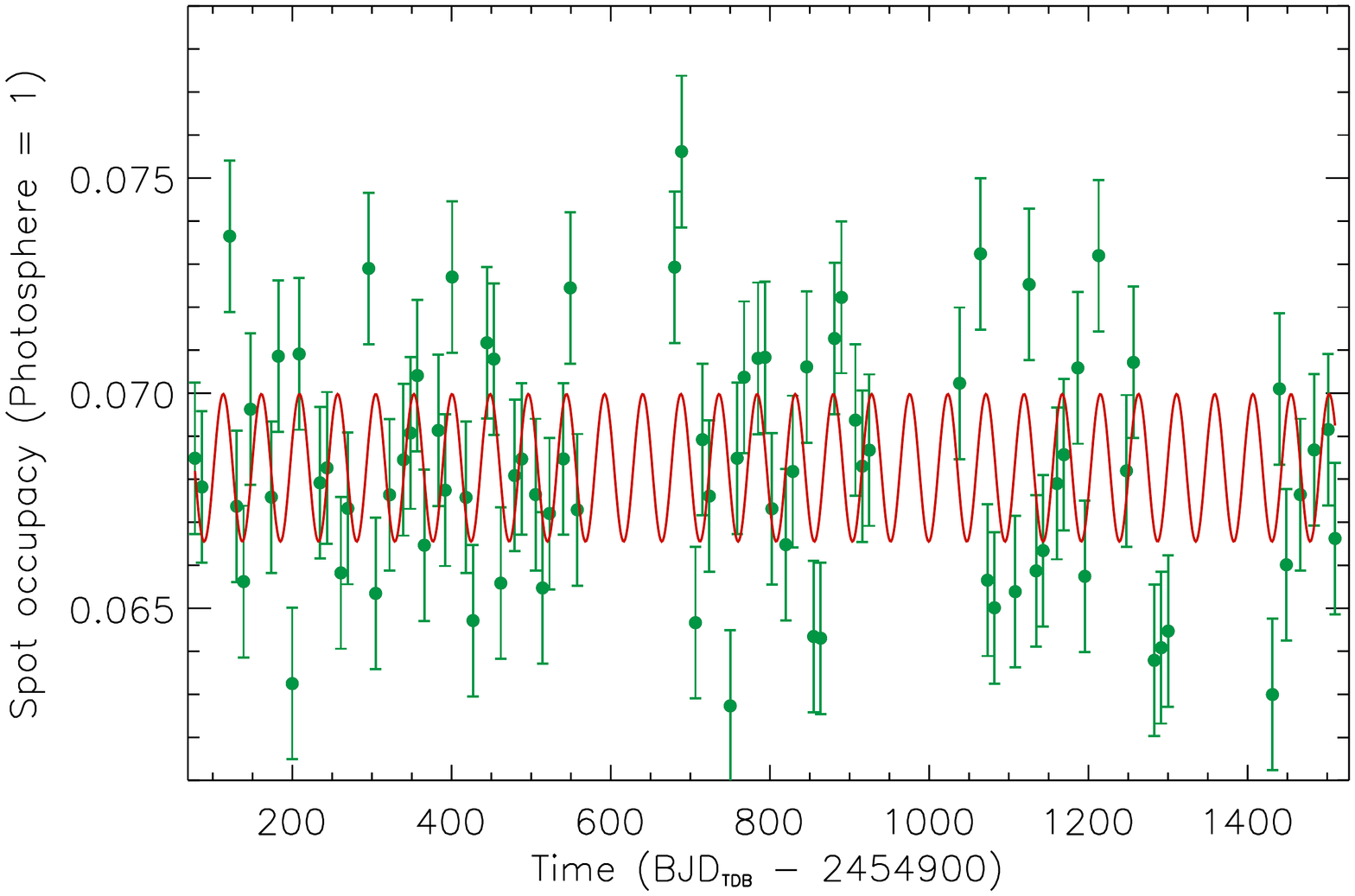}} 
   \caption{The total spotted area as derived from the ME best fits to the light curve de-trended with the method of \citet{BonomoLanza12} vs. the time (green filled circles).  The error bars have an amplitude of $3\sigma$, where $\sigma$ is the standard deviation as derived from the photometric accuracy of the datapoints. Values for the intervals with $\delta > 0.2$ have been excluded. The best fitting sinusoid with a period of 47.906 days is superposed to the time series (red solid line). }
              \label{fig10}%
\end{figure}
\begin{figure}
\hspace*{-1.5cm}
 \centering{
 \includegraphics[width=8cm,height=11cm,angle=90]{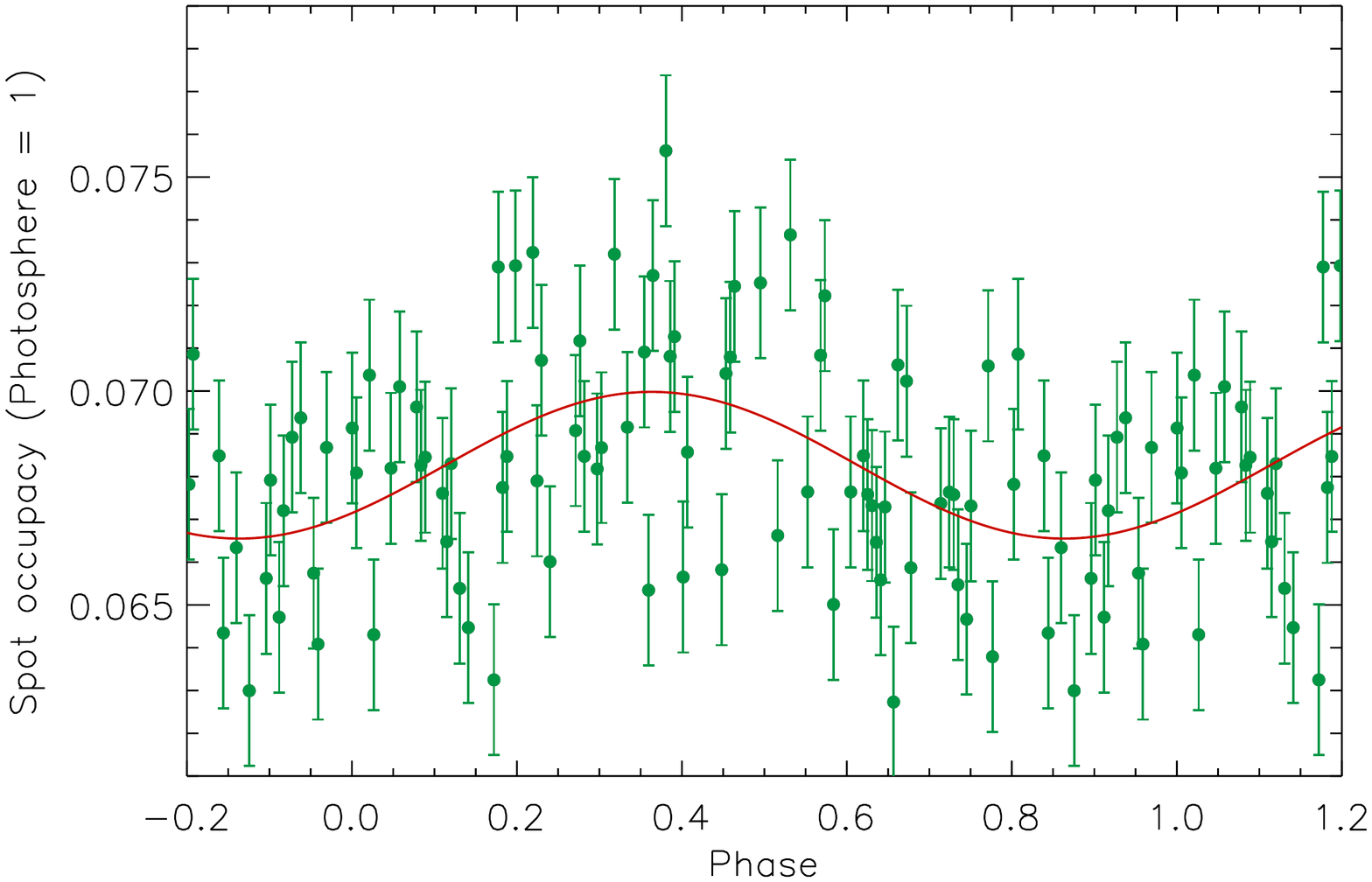}} 
   \caption{Same as Fig.~\ref{fig10}, but with the area values put in phase with the period of 47.906~days. The best fitting sinusoid with a period of 47.906 days is superposed to better show the oscillation of the spotted area (red solid line). }
              \label{fig10phased}%
\end{figure}

\begin{figure}
\hspace*{-1.5cm}
 \centering{
 \includegraphics[width=8cm,height=11cm,angle=90]{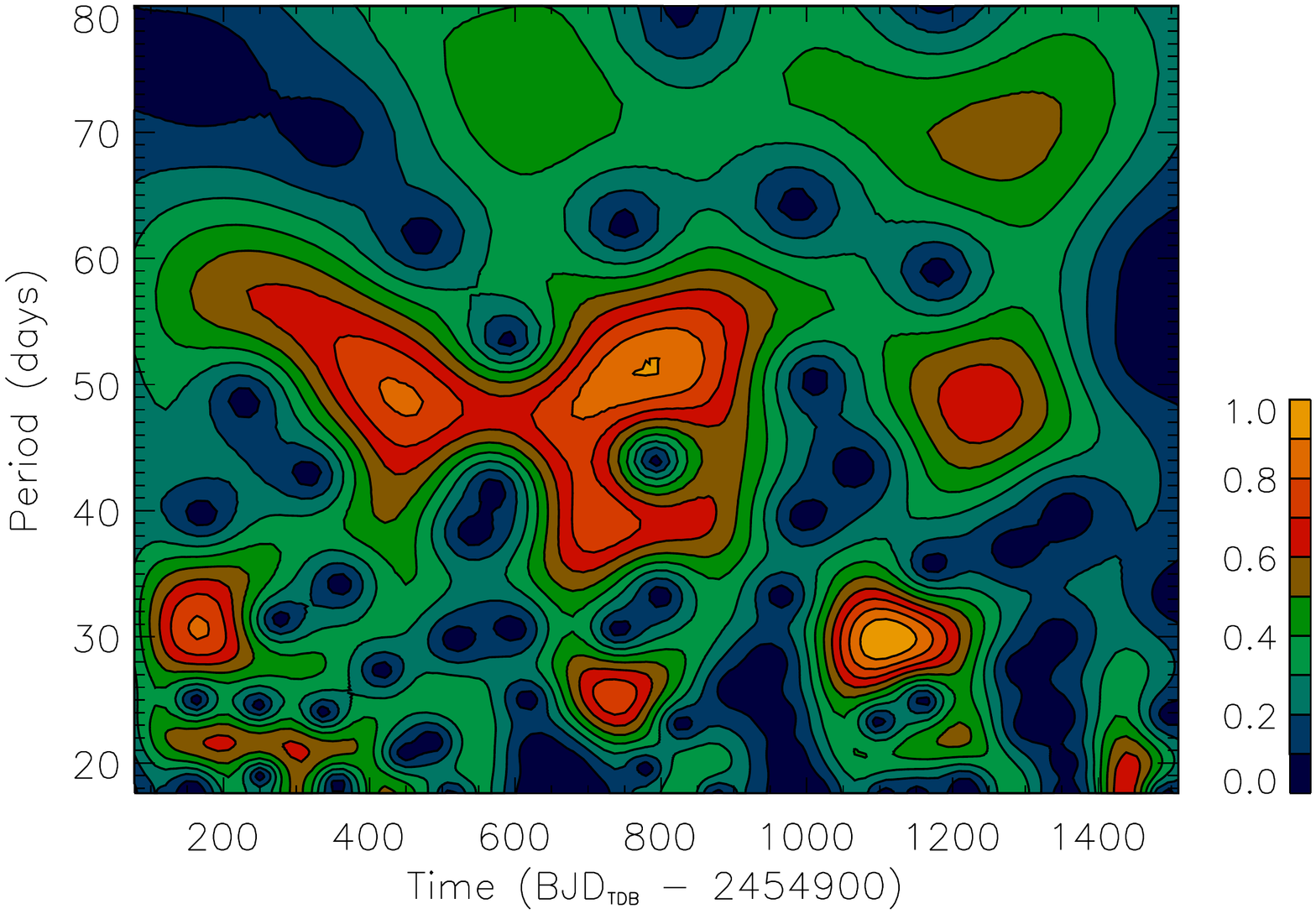}} 
   \caption{Amplitude of the Morlet wavelet of the total spotted area variation in Fig.~\ref{fig10} vs. the period and the time. The amplitude was normalized to its maximum value. Different colours indicate different relative amplitudes from the maximum (orange) to the minimum (dark blue) as indicated in the colour scale in the right lower corner.}
              \label{fig11}%
\end{figure}
\begin{figure}
\hspace*{-1.5cm}
 \centering{
 \includegraphics[width=8cm,height=11cm,angle=90]{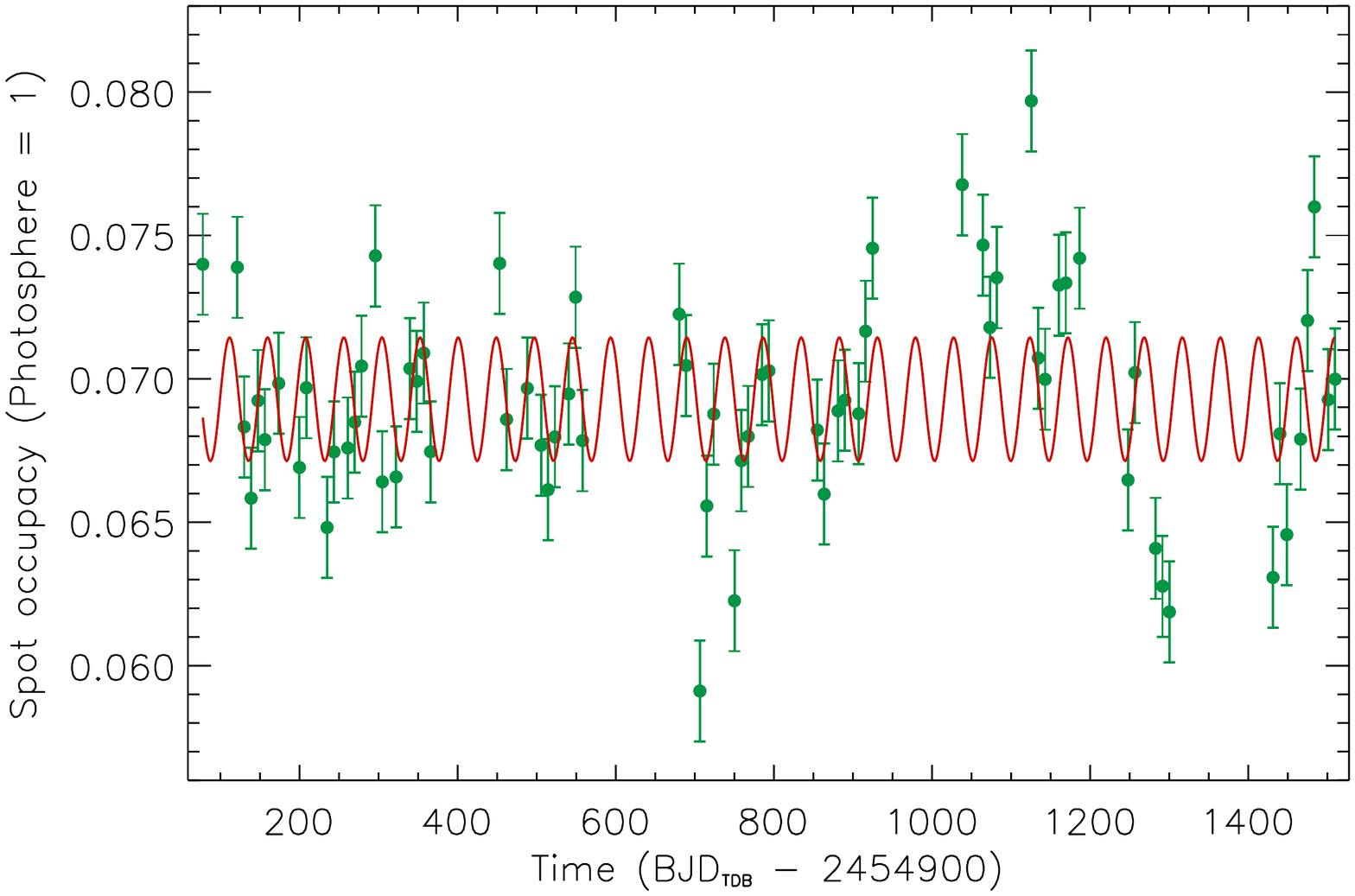}} 
   \caption{Same as Fig.~\ref{fig10}, but for the area values derived by the ME best fit to the ARC2 light curve. Intervals with $\delta > 0.25$ have been discarded. The period of the GLS best fitting sinusoid is 48.202 days. Note the different scale on the y-axis. }
              \label{fig12}%
\end{figure}
\begin{figure}
\hspace*{-1.5cm}
 \centering{
 \includegraphics[width=8cm,height=11cm,angle=90]{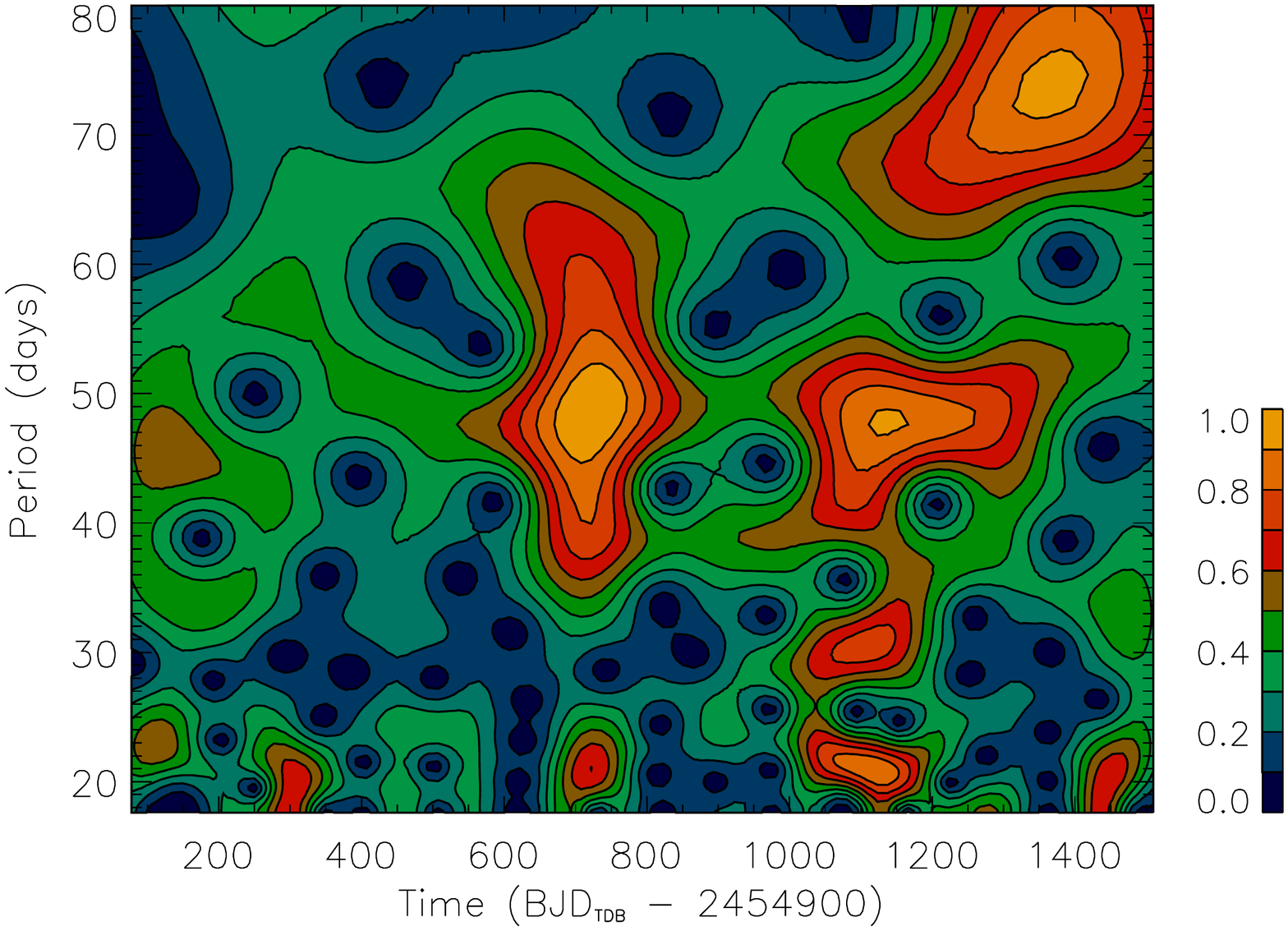}} 
   \caption{Same as Fig.~\ref{fig11}, but for the time series in Fig.~\ref{fig12}.}
              \label{fig13}%
\end{figure}

The variation of the total spotted area as derived from the intervals of the ARC2 light curves with $\delta \leq 0.25$ are plotted in Fig.~\ref{fig12} together with the GLS best fitting sinusoid. It has a period of 48.202~days, but its analytic FAP is 0.1667, while the FAP from $10\,000$ shuffling is 0.40, likely as a consequence of the lower number of datapoints and larger fluctuations from one interval to the next. The corresponding Morlet wavelet map is plotted in Fig.~\ref{fig13} and its overall aspect is similar to that of the map in Fig.~\ref{fig11}, reinforcing the case for a periodicity in the spotted area of $\sim 50$ days during the first half of the time series. In the second half, the wavelet power is split among several different periodicities, likely as a consequence of the gaps in the time series and a long-term modulation with a period of several hundred days appearing for $t^{\prime} \ga 800$~days. Such a modulation is not observed in the area time series obtained from the light curve de-trended with the approach of \citet{BonomoLanza12} that fits and removes a parabolic trend within each quarter, thus filtering out the variations on timescales comparable with the quarter duration of $\sim 90$~days (cf. the photometric time series in Figs.~\ref{fig1},~\ref{fig6}, and~\ref{fig4}).  

{Unfortunately, Kepler data are not useful to search for long-term changes of the mean light level of Kepler-17 to confirm the activity cycle of $\approx 400-600$~days suggested by the different regimes of spot longitude  migration (cf. Sect.~\ref{spot_active_longitudes}). Some hint of a long-term variation of the spotted area may be apparent in Fig.~\ref{fig12}, but the lack of a complete correction for the systematic variations from one quarter to the next hampers our attempts to confirm this result. Note that the variation of the spotted area plotted in Figs.~\ref{fig10} and~\ref{fig12} refers to the spots that are unevenly distributed in longitude, because the  amplitude of the rotational modulation is  insensitive to uniformly distributed spots. In other words, if the cycle of $\approx 400-600$~days is associated with a variation of the area of spots almost uniformly distributed in longitude, it can go undetected in those plots and only a long-term variation of the mean light level would reveal its presence. With a chromospheric index $\log R^{\prime}_{\rm HK} = -4.47$ \citep[][Sect.~4.2.2]{Bonomoetal12},  Kepler-17 is at the boundary separating very active stars with predominantly non-axisymmetric spot distributions and active longitudes from less active rotators with an almost uniform distribution of  spots in longitude. This may account for its complex behaviour probably showing phenomena common to both kinds of stars \citep[][]{Lehtinenetal16}. } 

\subsection{Star-planet interaction: tides}

We did not find evidence of a light modulation associated with a possible star-planet interaction in the residuals of our spot models (cf. Sect.~\ref{light_curve_models}). However, some effect of the planet on the star is expected  because of its mass of $\sim 2.5$ Jupiter masses and its proximity. Tides raised on the star by the planet are an example of such an interaction. {Thanks to our determination of stellar rotation, we can derive information on the  tidal dissipation inside the G2V star Kepler-17 that is useful to model the evolution of the rotation itself. This can be applied to evaluate the activity level of the star in the past, thus providing information for models of planetary evolution and evaporation \citep[e.g.][]{Murray-Clayetal09}, as well as the confidence of stellar age based on gyrochronology. }

The orbital angular momentum of the planet is $\sim 4$ times the stellar spin angular momentum, while the total angular momentum of the system is only $\sim 0.6$ of that required to reach a synchronous final state, in the hypothesis that the total angular momentum of  the system is conserved \citep{Hut80}. However, the stellar magnetized wind produces a steady loss of angular momentum from the system that accelerates the shrinking of the planetary orbit until the planet will be engulfed by the star  \citep[cf.][]{DamianiLanza15}. 

We investigate the evolution of the stellar rotation and the orbital semimajor axis by applying the simple model of \citet{LanzaMathis16} that includes the  wind braking of the stellar rotation using a Skumanich-type law. The star is assumed  to rotate rigidly and the strength of the tidal interaction is parameterized by the stellar modified tidal quality factor $Q^{\prime}$ \citep{Zahn08}. A stronger interaction implies a faster dissipation of the kinetic energy of the tides and is parameterized by a smaller value of $Q^{\prime}$. We include in the model the evolution of the radius of the star calculated by means of the EZweb interface\footnote{http://www.astro.wisc.edu/\~{}townsend/static.php?ref=ez-web} because the tidal torque is proportional to $(R/a)^{6}$, where $R$ is the stellar radius and $a$ the orbit semimajor axis \citep{Zahn08}. We consider a model for a main-sequence star of mass $1.095$~M$_{\odot}$ and metal abundance $Z= 0.03$ because it has a radius of $1.06$~R$_{\odot}$ at the estimated age of Kepler-17, that is 1.8~Gyr \citep[cf.][]{Bonomoetal12}. {This value of the mass is different from that  derived by \citet{Desertetal11} and \citet{Bonomoetal12} by fitting different stellar evolution models to the position of the star in the mean density-effective temperature diagram, but it is still within $\sim 1\sigma$ from their mass estimates. We prefer to adopt a stellar evolution model that fits the radius at the putative age of 1.8~Gyr rather than the estimated mass because the radius evolution has a much stronger impact on the tidal evolution of the system, while the ratio of the stellar mass to the planetary mass stays fixed in our model.} We assume a circular orbit because tides inside the planet damp any initial eccentricity on timescales of $10-100$~Myr, i.e., much shorter than the age of the star. The obliquity of the planetary orbit is assumed to be zero following the discussion in Sect.~7.2.1 of  \citet{Desertetal11}. 

The strength of the tidal interaction in star-planet systems is unknown and there are theoretical reasons to believe that it depends on the ratio between the tidal frequency\footnote{Considering the semidiurnal tide as the dominant component, the tidal frequency is $\hat{\omega}= 2(n-\Omega)$, where $n= 2\pi/P_{\rm orb}$ is the orbital frequency and $\Omega = 2\pi/P_{\rm rot}$ the spin frequency of the star.} and the rotation frequency of the star \citep{OgilvieLin07}. Observational estimates have been performed only with statistical methods that do not provide information on individual systems, but only an indication of the mean values of $Q^{\prime}$ in different regimes \citep[e.g.][]{Bonomoetal17,CollierCameronJardine18}.  Therefore, we model the evolution assuming two constant values for $Q^{\prime}$, that is $10^{7}$ and $10^{8}$, with a preference for the latter from a theoretical point of view. Specifically, the tidal frequency and the rotation frequency in Kepler-17 are always sufficiently apart as to avoid the excitation of inertial waves,  {that is, $\hat{\omega} > 2\Omega$,} thus leading to a weaker tidal interaction between the star and the planet \citep[cf.][]{OgilvieLin07}. {Conversely in the $|\hat{\omega}| \leq 2\Omega$ regime, the excitation of inertial waves, strongly dissipated in wave attractors, would lead to a decrease of $Q^{\prime}$ by $2-3$ orders of magnitudes \citep[cf.][]{Rieutordetal01,OgilvieLin07,GoodmanLackner09}. }

In Fig.~\ref{fig15} we plot the evolution of the stellar spin and orbit semimajor axis together with the evolution of the stellar radius adopted to compute the first two quantities. We also plot the evolution of the stellar rotation without any tidal torque, that is under the action of the Skumanich-type wind braking only. We assume that our model can be applied for ages later than $\sim 0.6$~Gyr because younger ages may still show the effects of the initial conditions and of an incomplete internal core-envelope coupling.  The stellar age is assumed to be 1.8~Gyr and the presently measured mean rotation period of 12~days is imposed at that age to all our models.  

The spin evolution is dominated by magnetic braking up to about $\sim 2$~Gyr for $Q^{\prime} = 10^{8}$, while it deviates remarkably from the Skumanich law for $Q^{\prime}=10^{7}$ all along the evolution {because the angular momentum exchange due to tides dominates over the loss of angular momentum by the wind braking. This leads to a remarkably longer rotation period of the star in the past than in the case with $Q^{\prime} = 10^{8}$ because  the planet spun up the star through its tidal interaction, while its orbit was decaying.} However, even for $Q^{\prime}=10^{8}$, the slope of the braking law is reduced by the tidal interaction with tides that counteract magnetic braking leading to a less steep variation of the rotation period. This suggests that the age estimated by means of standard gyrochronology, {that ranges between 1.0 and 1.4~Gyr \citep{Barnes10,Barnesetal16}},  is not accurate for Kepler-17 because of the tidal spin-up induced by its massive and close-by planet. Moreover, the planet could also affect the efficiency of the stellar wind \citep[cf.][]{Cohenetal10,Lanza10}. {Therefore, the present results support the adoption of a modified gyrochronology relationship to evaluate the age of Kepler-17 \citep[cf.][]{Lanza10} and provide a rotation evolution scenario in agreement with the age of 1.8~Gyr as estimated by \citet{Bonomoetal12}.}

The expected  survival time of the planet  is estimated as $\approx 0.35$~Gyr in the case of the stronger tidal interaction and $\approx 2.4$~Gyr in the case of the weaker interaction. The orbital decay is mainly ruled by the increase of the stellar radius along its main-sequence evolution owing to the remarkable dependence of the tidal torque on $R/a$. 
\begin{figure}
 \centering{
 \includegraphics[width=9cm,height=10cm,angle=0]{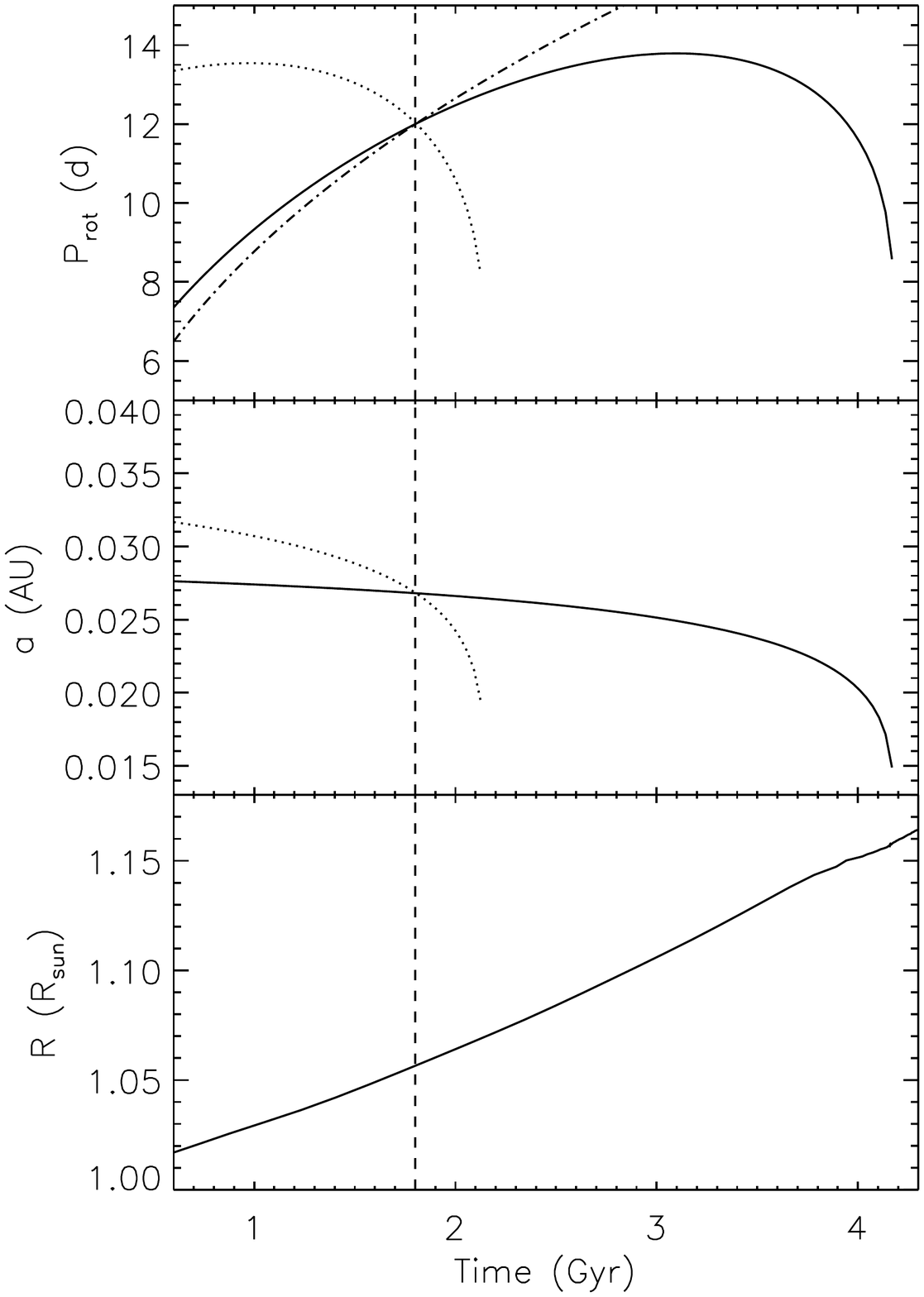}} 
    \caption{Top panel: The mean rotation period of Kepler-17 vs. the time for our  tidal evolution model computed with $Q^{\prime} = 10^{7}$ (dotted line) and $Q^{\prime}=10^{8}$ (solid line). The evolution of the stellar spin without any tidal torque and assuming a period of 12.01 days at an age of 1.8~Gyr is plotted as the dot-dashed line. The vertical dashed line indicates the adopted age of the star. Middle panel: same as the top panel, but for the orbit semimajor axis. Lower panel: the radius of the star  adopted to compute the tidal evolution vs. the time. }
              \label{fig15}%
\end{figure}

\section{Discussion and conclusions}
\label{discussion}

{We have analysed the activity of Kepler-17 using two different approaches to de-trend the systematics present in Kepler timeseries. We confirm that the PDC pipeline introduces an overcorrection of the subtle light modulations produced by solar-like faculae in late-type stars making PDC timeseries not recommendable to produce spot maps by light curve inversions \citep[see also][]{Aigrainetal17}. Conversely, the de-trending by the ARC2 pipeline suffers from much less problems and gives results comparable with those derived by  the simpler approach by \citet{BonomoLanza12}, at least in the case of this target with high signal-to-noise ratio data and high level of activity. Nevertheless, the evaluation of the total spotted area from the ARC2 timeseries can still be affected by some residual trends on timescales ranging from $\sim 10$ to $\sim 90$~days, probably due to the use of a limited number of CBVs to correct the light curve. }

Our results provide an extended comparison of the maps obtained from the in-transit and out-of-transit light modulations, that is from two independent datasets and methods. The good correspondence found during certain  time intervals indicates that the adopted maximum-entropy approach is capable of reconstructing the overall starspot  distribution in longitude and time and gives support to the existence of active longitudes. 
A similar comparison was made by \citet{SilvaValioLanza11} validating the spot models for CoRoT-2, but it was based only on $\sim 150$~days of data, while the present comparison is more extended and shows the effects of a likely activity cycle of $\approx 400-600$~days. 

The active longitude around $\sim 200^{\circ}$ is remarkable because it lasts for at least 1400 days, although its level of activity is continuously changing  as indicated by the varying spot filling factors. Similar long-lived active longitudes are commonly observed on very active rotators such as young solar-type stars \citep{Lehtinenetal16} or the subgiant members of the close active binaries II~Peg \citep{Rodonoetal00} or HR~1099 \citep{Lanzaetal06}. Space-borne photometry has revealed an active-longitude phenomenology similar to that of Kepler-17 in, e.g., CoRoT-2 \citep{Lanzaetal09} or CoRoT-6 \citep{Lanzaetal11} or in the M dwarf GJ 1243 \citep{Davenportetal15}. Active longitudes have been proposed also for the Sun, although they are not as evident as in more active and rapidly rotating  solar-like stars \citep[e.g.,][]{Usoskinetal07}. Recent hydromagnetic dynamo models have provided some insight on the physical mechanisms that could produce such active longitudes \citep{Weberetal13}. 

We found some indication of an activity cycle of $\approx 400-600$~days from the latitudinal migration of the starspots in Kepler-17. {\citet{EstrelaValio16} found indication for a similar periodicity in the area of the spots occulted during transits.} This cycle could be similar to the solar eleven-year cycle, although of remarkably shorter duration. Short activity cycles have recently been found in $\iota$~Horologii, an F8V star that hosts a giant planet on a 300-d orbit and shows a cycle of $\sim 585$~days \citep{SanzForcadaetal13}; and in the young ($\sim 1$~Gyr) G1.5V~star HD~30495 that has a rotation period of $\sim 11$~days and  shows two chromospheric cycles, one of $620 \pm 150$~days and another of $12 \pm 3$~years. The short-term modulation is intermittent and does not appear to be related to the longer-term cycle \citep{Egelandetal15}. The case of HD~30495, that is similar to Kepler-17 in effective temperature and rotation period, suggests that such short cycles may be a characteristic of young Sun-like stars. Other possible examples are HD~76151, a G3V star with a rotation period of $\sim 15$~days and a cycle of $\sim 920 \pm 10$~days;  and HD~190406, a G1V star with a rotation period of $\sim 14$~days and two cycles of $\sim 950 \pm 10$~days and $\sim 17$~years \citep[see][]{Baliunasetal95,Baliunasetal96}. 

In addition to this  possible activity cycle in Kepler-17, we find {marginal evidence for an oscillation of the total spotted area with a period of $\sim 48$~days, previously reported by \citet{BonomoLanza12}.} {The mean total area of the occulted spots found by \citet{Valioetal17} is $\sim 6 \pm 4$ percent that agrees with our mean spotted area. However, they do not find evidence of  the 48-days periodicity probably because the large variations in the area of the individual spots hamper its detection making clearly apparent only the periodicity at the mean rotation period \citep[cf. Fig.~6 in][]{Valioetal17}.} The 48-days spotted area modulation is not associated with a migration of the main latitude of spot formation and is reminiscent of the so-called Rieger cycles in the Sun \citep[e.g.][]{Oliveretal98,Zaqarashvilietal10,Gurgenashvilietal17}. They have been attributed to Rossby-type waves propagating in the solar interior that modulate the toroidal magnetic field responsible for the formation of the spots \citep[cf.][]{Zaqarashvilietal10,Gurgenashvilietal16,Zaqarashvili18}. Similar cycles  have been observed in, e.g., CoRoT-2 \citep{Lanzaetal09} and in some young late-type stars with an age between 4 and 95~Myr investigated by \citet{Distefanoetal17}. Short-term cycles, possibly of Rieger type, have also been investigated using CoRoT \citep{FerreiraLopesetal15} and Kepler \citep{Arkhypovetal15} time series. 

The amplitude of the latitudinal differential rotation in Kepler-17 derived from our spot modelling is only a lower limit because we do not know the latitudes of the spots rotating with different periods. Moreover, the rotation of the {overall spot pattern} can be different from that of individual starspots as suggested in the case of CoRoT-2 by \citet{Froehlichetal09} (cf. also Appendix~\ref{app2}). Therefore, our estimate of $\sim 14 \pm 5$ percent relative amplitude of the differential rotation should be taken with some caution. {Moreover, this amplitude is reduced to about $8 \pm 5$ percent for some values of our spot modelling parameters (cf. Appendix~\ref{app1}).} Nevertheless, the solar-like character of the differential rotation, i.e., the faster rotation of the equator with respect to the higher latitudes, is well established, thanks to the comparison with the spots occulted during the transits that are certainly located at low latitudes.  Using the measurements of the rotation periods of those spots and assuming a solar-like differential rotation profile, \citet{Valioetal17} estimated a relative pole-equator angular velocity difference $\Delta \Omega /\Omega \simeq 8.0 \pm 0.9$ percent, {close to the lower limit of our determination. }

The range of rotation periods derived from the modulation of the chromospheric flux in late-type stars by \citet{Donahueetal96} suggests $\Delta P_{\rm rot}/P_{\rm rot} \sim 0.13$ for Kepler-17, although HD~190406, that has similar spectral type and mean rotation period, shows $\Delta P_{\rm rot}/P_{\rm rot} = 0.21$. The large statistical sample considered by \citet{ReinholdGizon15} shows a relative amplitude up to $0.1-0.2$ for G-type stars with the mean rotation period of Kepler-17, based on the analysis of the photometric time series of Kepler targets. 
Recent theoretical models by \citet{Brunetal17} predict a solar-like differential rotation for Kepler-17, that is with the equator rotating faster than the poles. Its fluid Rossby number \citep[see][eq.~33 and Fig.~22]{Brunetal17} is $R_{\rm of} \sim 0.6$ giving an expected relative amplitude of the differential rotation between the equator and $60^{\circ}$ latitude of $\Delta \Omega / \Omega \simeq 0.2$. Therefore, we conclude that the amplitude of the latitudinal differential rotation of Kepler-17 as derived from our analysis is in agreement with both observations and theoretical models for stars of similar spectral type and mean rotation period. 

In the Kepler-17 system, the tidal interaction between the planet and the star is likely to be relevant. We find that it is capable of modifying the evolution of the stellar rotation by counteracting the braking by the stellar wind. Even if we assume a weak tidal coupling ($Q^{\prime} \sim 10^{8}$), that is favoured in our model,  the stellar spin up is significant and makes it impossible to derive a precise age of the star by means of gyrochronology \citep{Barnes07,Barnes10}. This is probably the case of several stars hosting massive close-by planets as discussed by, e.g., \citet{Ferraz-Melloetal15} and \citet{DamianiLanza15}. {Nevertheless, an estimated age of $\sim 1.8$~Gyr \citep{Bonomoetal12} is in agreement with a simple model of the tidal evolution of the system. The same model predicts a rotation period of the star of $\sim 7.5$~days at an age of 0.6~Gyr and an almost constant orbit semimajor axis for the planet over the $0.6-1.8$~Gyr time interval. This information can be used to evaluate the evolution of the stellar high-energy radiation flux \citep[e.g.,][]{Sanz-Forcadaetal11}, that affects the evaporation of the planet, and is controlled by the rotation rate of the star itself \citep[e.g.][]{Schmitt10}.  }

\begin{acknowledgements}
The authors are grateful to an anonymous referee for a careful reading of their manuscript and valuable comments that improved the presentation of their results. 
AFL is grateful for interesting discussions with the collegues of the team on "Rossby waves in Astrophysics" led by Prof.~T.~Zaqa\-rashvili and supported by the International Space Science Institute in Bern, Switzerland. AFL and ASB acknowledge support from INAF/Frontiera through the Progetti Premiali funding scheme of Italian Ministry of Education, University, and Research. YN gratefully acknowledges support for the Programa de Doutorado Sandu\'{\i}che no Exterior (PDSE) from the Brasilian Federal Agency for Support and Evaluation of Graduate Education (CAPES) during his stage at INAF-Catania Astrophysical Observatory (Process no. 88881.134871/2016-01).
\end{acknowledgements}

\appendix
\section{Effects of changing  model parameters on the spot distributions}
\label{app1}
{We explore the effect of varying our model parameters on our main results. Specifically, we assume theoretical limb-darkening coefficients in place of those derived from the fitting of  the transits by \citet{Maxted18}; or  change the contrast of the spots $c_{\rm s}$; or vary the facular-to-spotted area ratio $Q$. In this investigation, we change one parameter at a time to isolate its effects on our results. We present results for the light curve detrended according to the method of Bonomo \& Lanza because it has less gaps than the ARC2 light curve and because the results are very similar. We focus on the migration rate of the spots and on the variation in their total coverage that affect our measurements of the differential rotation and of the activity level, respectively. 

\subsection{Varying the limb-darkening coefficients}
In Fig.~\ref{lag_migration_LDC}, we plot the migration rate vs. the time as derived from the ME spot modelling with the limb-darkening coefficients obtained from the fitting of the transits according to \citet{Maxted18} (green diamonds) or from model atmospheres  \citep[orange triangles; see][and discussion in Sect.~\ref{parameters}]{Muelleretal13}. The migration rates of the spot distributions are closely comparable, if we consider only the values between $-4$ and $+2$ deg/day. Therefore, the amplitude and the sign of the differential rotation are not affected, if we adopt the theoretical limb-darkening coefficients. 

The total spotted area shows a systematic difference of about $ -2.8$~percent with oscillations not exceeding $0.4$~percent that do not affect our conclusions concerning possible activity cycles in Kepler-17 (cf. Fig.~\ref{area_comparison_LDC}). Specifically, the GLS periodogram of the modulation of the spotted area has its maximum at a period of 47.916~days, very close to that of 47.906 days obtained with our reference model; also the false-alarm probability is similar. 
\begin{figure}
\hspace*{-1.cm}
 \centering{
 \includegraphics[width=8cm,height=10cm,angle=90]{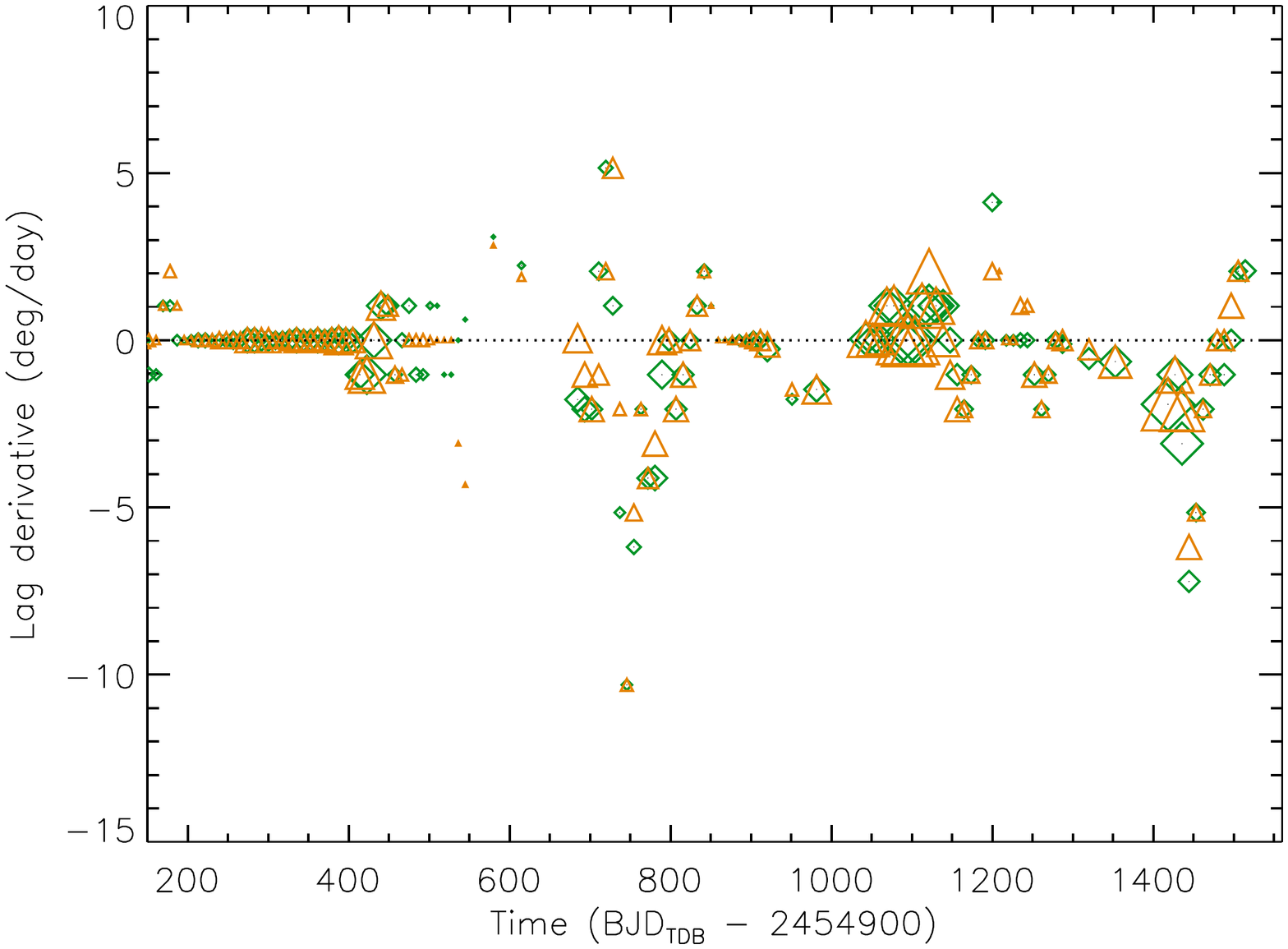}} 
\vspace*{-1.cm}
  \caption{Migration rate between consecutive spot pattern distributions as derived from the ME models of the Bonomo \& Lanza light curve  with the limb-darkening coefficients as derived by fitting the transits (green diamonds) or from model atmospheres (orange triangles; see our Sect.~\ref{parameters}). The size of the symbols is proportional to the cross-correlation coefficient $\rho_{\rm cc}$ (cf. Eq.~\ref{cc_eq}). }
              \label{lag_migration_LDC}%
\end{figure}
\begin{figure}
\hspace*{-1.cm}
 \centering{
 \includegraphics[width=8cm,height=10cm,angle=90]{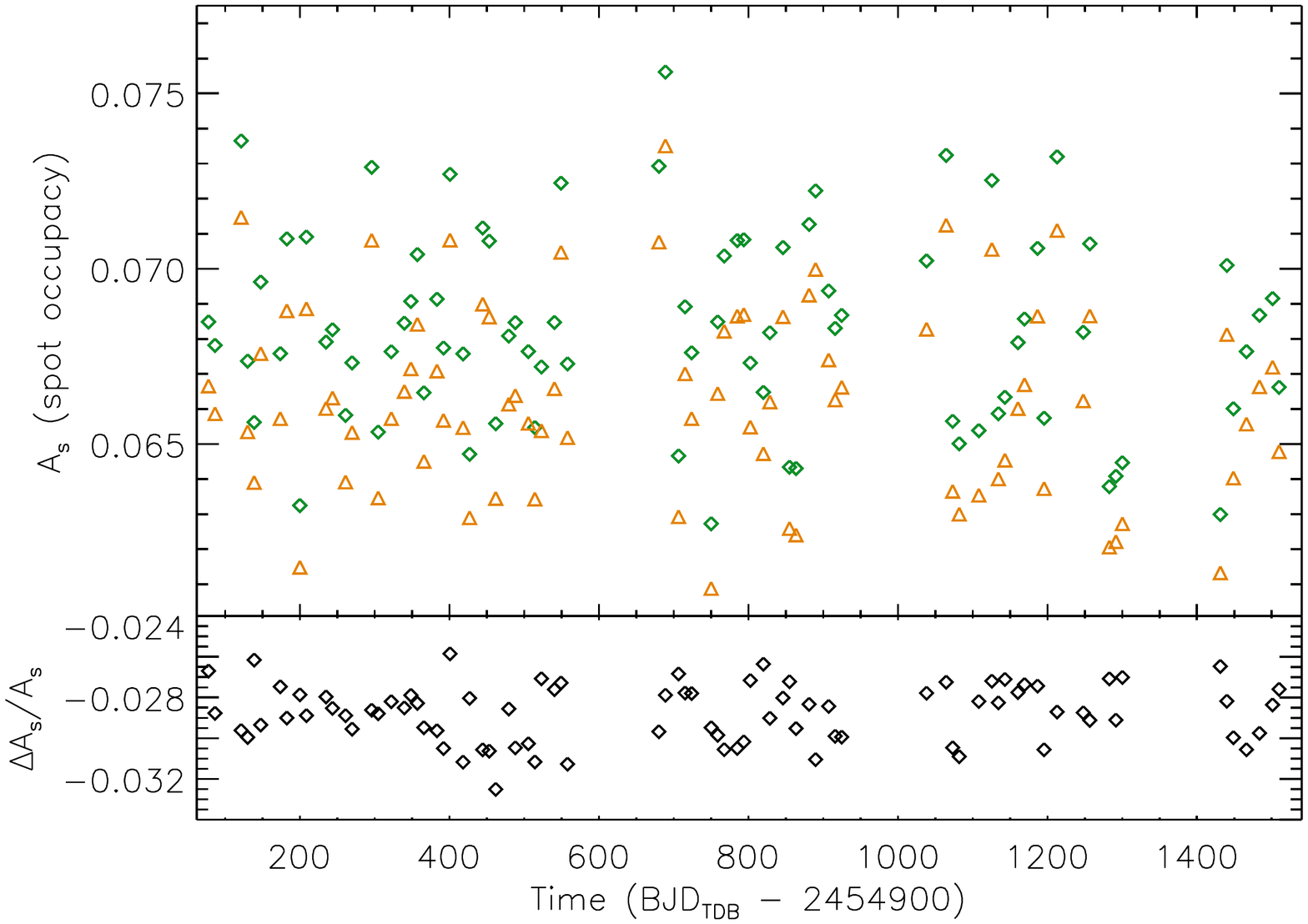}} 
\vspace*{-1.cm}
  \caption{Upper panel: Total coverage factor $A_{\rm s}$ of the starspots as derived from the ME models of the light curve of Bonomo \& Lanza with the limb-darkening coefficients as derived by fitting the transits (green diamonds) or from model atmospheres (orange triangles). Lower panel: relative difference between the values of the area obtained with the two sets of limb-darkening coefficients.}
              \label{area_comparison_LDC}%
\end{figure}

\subsection{Varying the spot contrast}
We explore the effect of varying the spot contrast between the extreme values measured by \citet{Valioetal17} by modelling spot occultations, that is, $c_{\rm s}\equiv I_{\rm spot}/I=0.38$ and 0.72; for comparison, in the case of sunspot groups, $c_{\rm s}=0.67$.  In Fig.~\ref{lag_migration_cs0.38}, we see that the spot migration rate has a less negative minimum value of about $ -2$ deg/day for $c_{\rm s}=0.38$, but a slightly greater positive maximum value, leading to a smaller amplitude of the relative differential rotation, that is, $\Delta P_{\rm rot}/P_{\rm rot} \sim 0.08 \pm 0.05$ when considering a period of 11.90 days for the rotation of  the occulted spots (cf. Sect.~\ref{spot_active_longitudes}). On the other hand, we find $\Delta P_{\rm rot}/P_{\rm rot} \sim 0.12 \pm 0.05$ when we consider $c_{\rm s}=0.72$ because a minimum migration rate of about $-3$ deg/day is measured during several cross-correlations (cf. Fig.~\ref{lag_migration_cs0.72}). 

The total spot coverage changes in a systematic way showing a smaller area  and a smaller amplitude of its modulation when the spots are darker, that is for $c_{\rm s}=0.38$ (cf. Figs.~\ref{area_comparison_cs0.38} and \ref{area_comparison_cs0.72}). However, the relative variations in the individual values of the coverage of $\pm\, 2-3$~percent  do not affect the main peak of the GLS periodogram that is at 47.916 and 47.962 days for $c_{\rm s} = 0.38$ and $0.72$, respectively. Only the value of the false-alarm probability as given by the formula of \citet{ZechmeisterKuerster09} is increased to $0.105$ in the case of $c_{\rm s} = 0.72$.   
\begin{figure}
\hspace*{-1.cm}
 \centering{
 \includegraphics[width=8cm,height=10cm,angle=90]{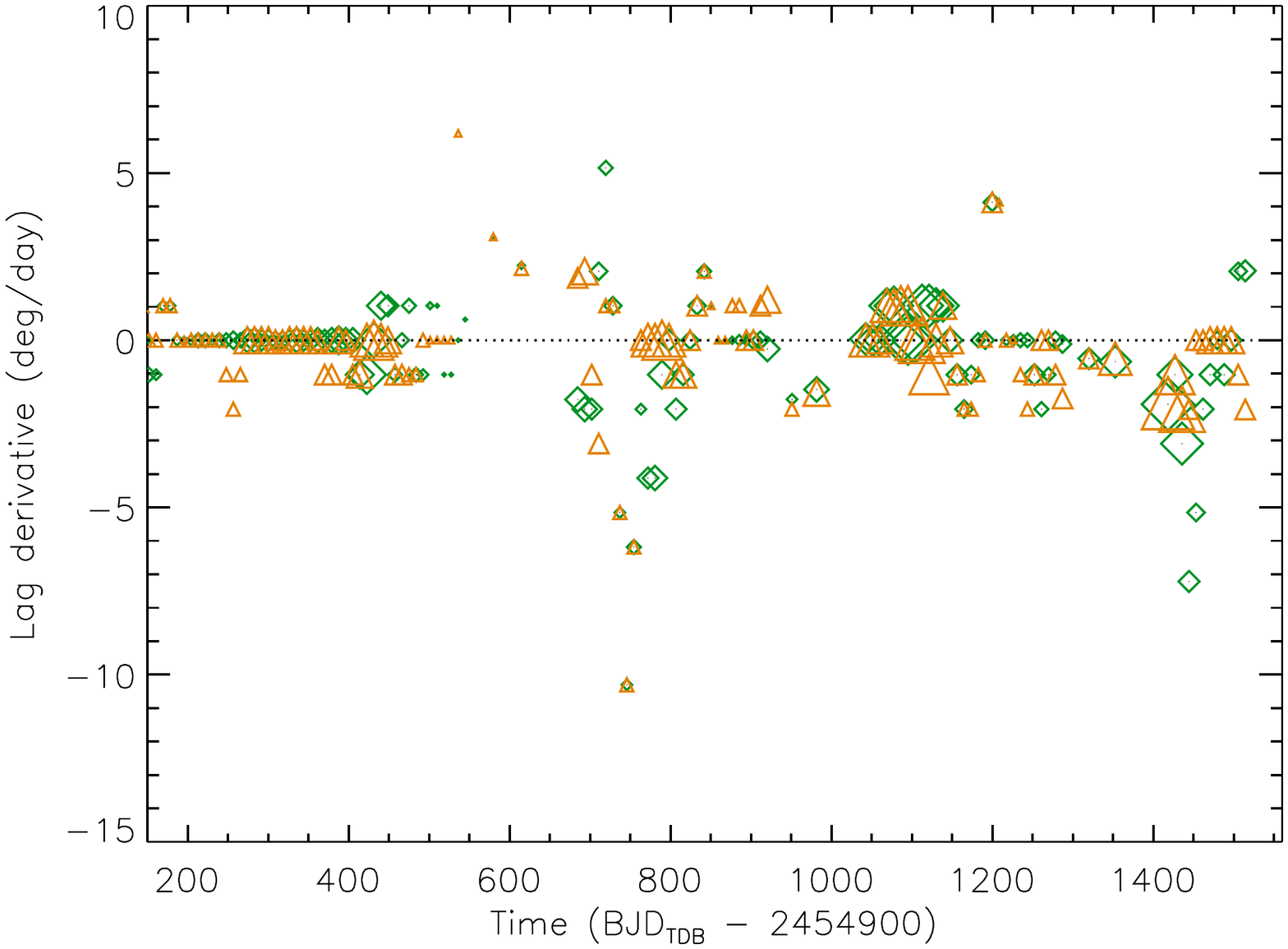}} 
 \vspace*{-1.cm}
  \caption{Migration rate between consecutive spot pattern distributions as derived from the ME models of the Bonomo \& Lanza light curve  with a spot contrast $c_{\rm s} = 0.55$  (green diamonds) or  $c_{\rm s}=0.38$ (orange triangles; see our Sect.~\ref{parameters}). The size of the symbols is proportional to the cross-correlation coefficient $\rho_{\rm cc}$ (cf. Eq.~\ref{cc_eq}). }
              \label{lag_migration_cs0.38}%
\end{figure}
\begin{figure}
\hspace*{-1.cm}
 \centering{
 \includegraphics[width=8cm,height=10cm,angle=90]{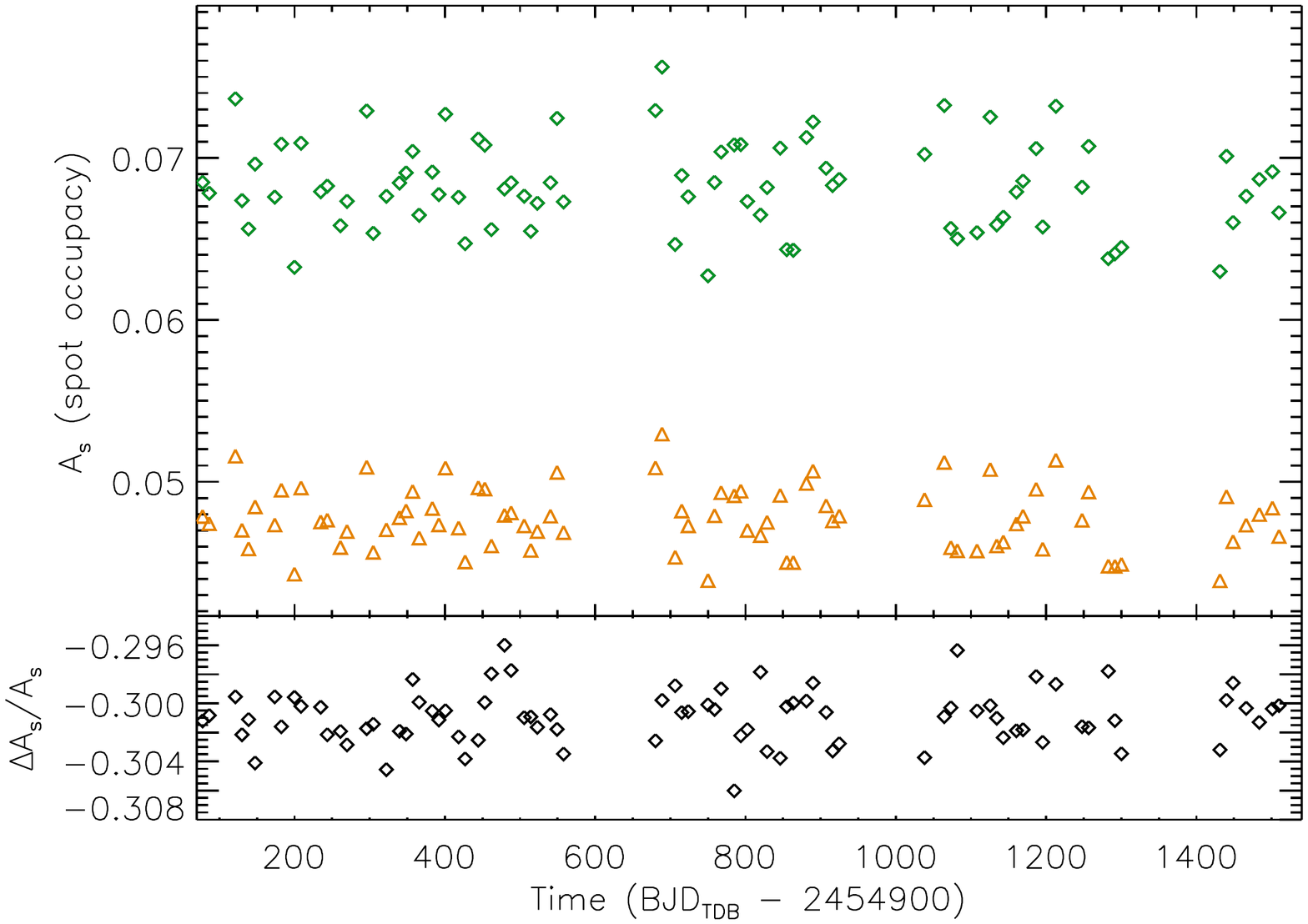}} 
\vspace*{-1.cm}
  \caption{Upper panel: Total coverage factor $A_{\rm s}$ of the starspots as derived from the ME models of the light curve of Bonomo \& Lanza  with $c_{\rm s}=0.55$ (green diamonds)  or $c_{\rm s}=0.38$ (orange triangles). Lower panel: relative difference between the values of the area obtained with the two different spot contrasts.}
              \label{area_comparison_cs0.38}%
\end{figure}
\begin{figure}
\hspace*{-1.cm}
 \centering{
 \includegraphics[width=8cm,height=10cm,angle=90]{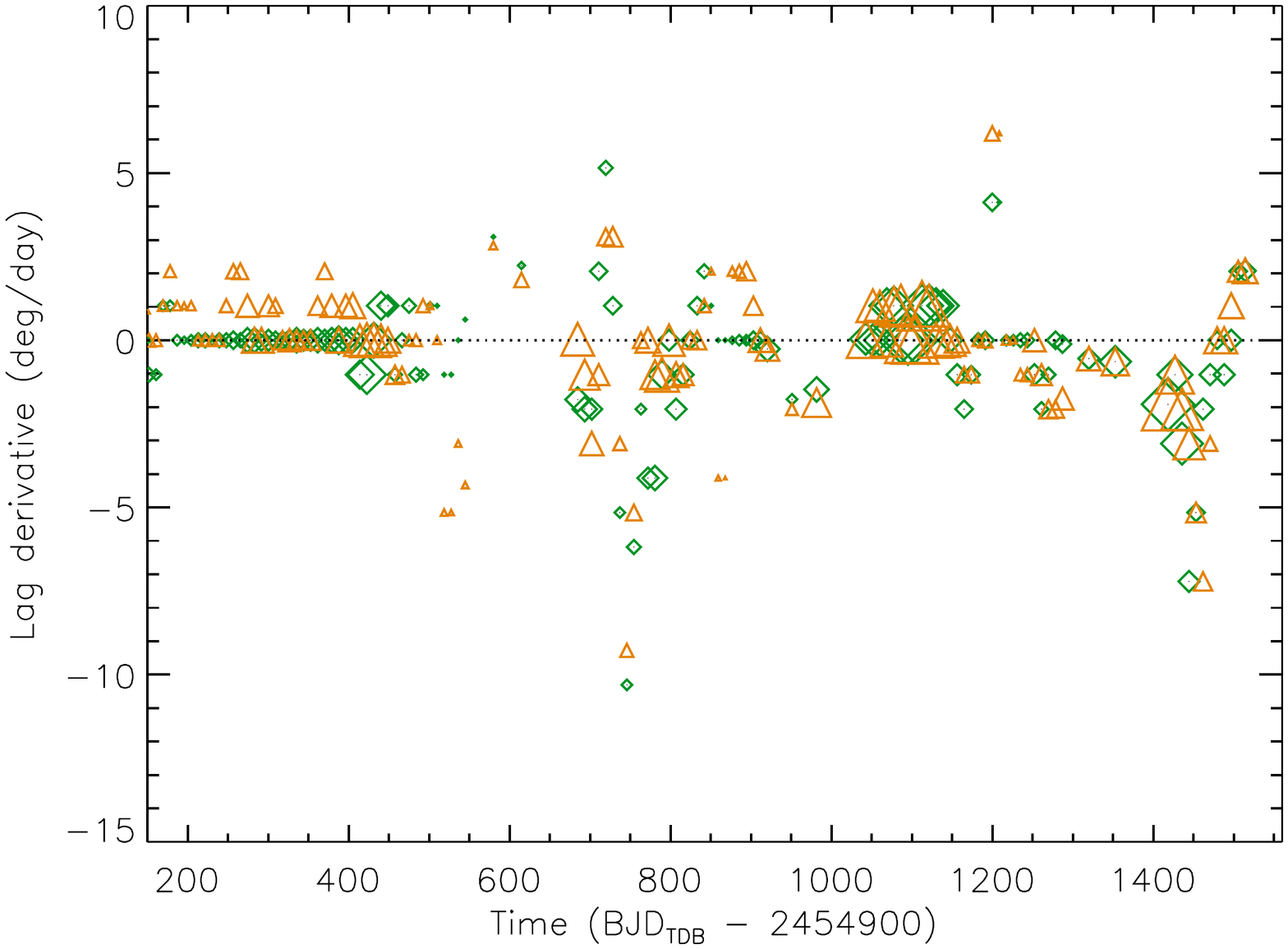}} 
\vspace*{-1.cm}
  \caption{Migration rate between consecutive spot pattern distributions as derived from the ME models of the Bonomo \& Lanza light curve  with a spot contrast $c_{\rm s} = 0.55$  (green diamonds) or  $c_{\rm s}=0.72$ (orange triangles; see our Sect.~\ref{parameters}). The size of the symbols is proportional to the cross-correlation coefficient $\rho_{\rm cc}$ (cf. Eq.~\ref{cc_eq}). }
              \label{lag_migration_cs0.72}%
\end{figure}
\begin{figure}
\hspace*{-1.cm}
 \centering{
 \includegraphics[width=8cm,height=10cm,angle=90]{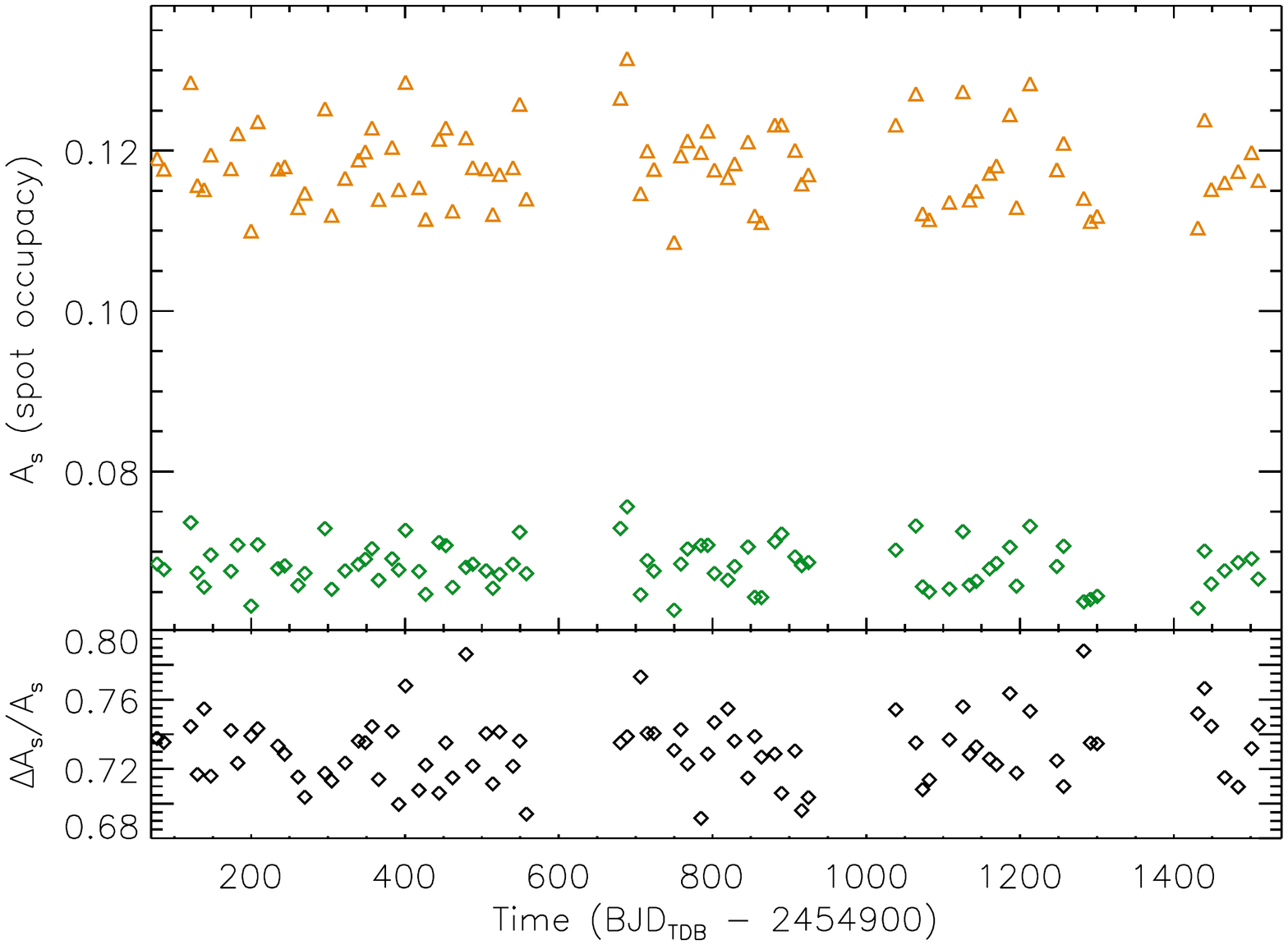}} 
\vspace*{-1.cm}
  \caption{Upper panel: Total coverage factor $A_{\rm s}$ of the starspots as derived from the ME models of the light curve of Bonomo \& Lanza with $c_{\rm s}=0.55$ (green diamonds)  or $c_{\rm s}=0.72$ (orange triangles). Lower panel: relative difference between the values of the area obtained with the two different spot contrasts.}
              \label{area_comparison_cs0.72}%
\end{figure}

\subsection{Varying the facular-to-spotted area ratio}
Finally, we explore the effects of the variation of the facular-to-spotted area ratio $Q$ between the extreme values $1.0$ and $4.0$ that are well beyond the 95~percent joint confidence interval of this parameter as derived in Sect.~\ref{parameters} by the analysis of the ARC2 and of the Bonomo \& Lanza light curves. The spot migration rate shows minimum negative values of about $-2$~deg/day for $Q=1.0$ (cf. Fig.~\ref{lag_migration_Q1}), while for $Q=4.0$, the minimum is about $-3$~deg/day (cf. Fig.~\ref{lag_migration_Q4}), corresponding to relative amplitudes of the differential rotation $\Delta P_{\rm rot}/P_{\rm rot} = 0.08 \pm 0.05$ and $0.12 \pm 0.05$, respectively. The variation of $Q$ produces systematic changes in the longitudes of the model active regions because the relative contributions of dark spots and bright faculae depend on their positions with respect to the centre of the stellar disc at a given rotation phase \citep[see][for a comparison of spot models with different $Q$'s with the position of sunspot groups]{Lanzaetal07}. 

The total spot coverage is affected by $Q$ because a larger spot area is required to counterbalance the effect of the larger faculae and reproduce the amplitude of the observed light modulation when $Q$ is increased (cf. Figs.~\ref{area_comparison_Q1} and~\ref{area_comparison_Q4}). In addition to this systematic variation, there are also fluctuations of relative amplitude of about $\pm\, 2$ percent with respect to the reference case with $Q=2.4$ that do not affect our results on a possible short-term activity cycle. Specifically, we find the maximum of the GLS periodogram of the area time series at periods of 47.955 and 47.962 days for $Q=1$ and $Q=4$, respectively. Only the analytic false-alarm probability is increased to $0.06$ and $0.12$, respectively, likely as a consequence of the non-optimal values of $Q$ adopted in those models. 
\begin{figure}
\hspace*{-1.cm}
 \centering{
 \includegraphics[width=8cm,height=10cm,angle=90]{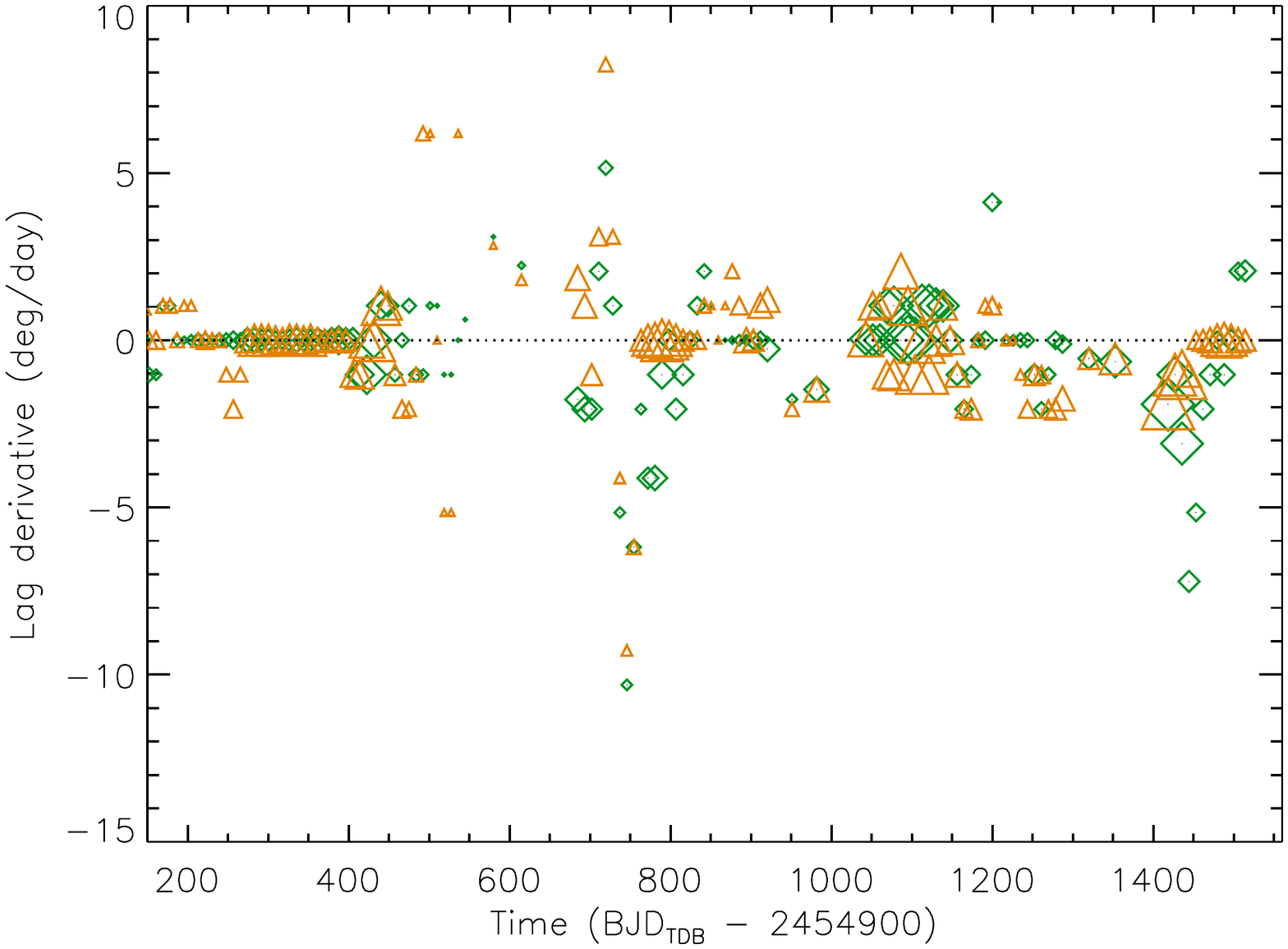}} 
\vspace*{-1.cm}
  \caption{Migration rate between consecutive spot pattern distributions as derived from the ME models of the Bonomo \& Lanza light curve  with facular-to-spotted area ratio $Q=2.4$  (green diamonds) or  $Q=1.0$ (orange triangles; see our Sect.~\ref{parameters}). The size of the symbols is proportional to the cross-correlation coefficient $\rho_{\rm cc}$ (cf. Eq.~\ref{cc_eq}). }
              \label{lag_migration_Q1}%
\end{figure}
\begin{figure}
\hspace*{-1.cm}
 \centering{
 \includegraphics[width=8cm,height=10cm,angle=90]{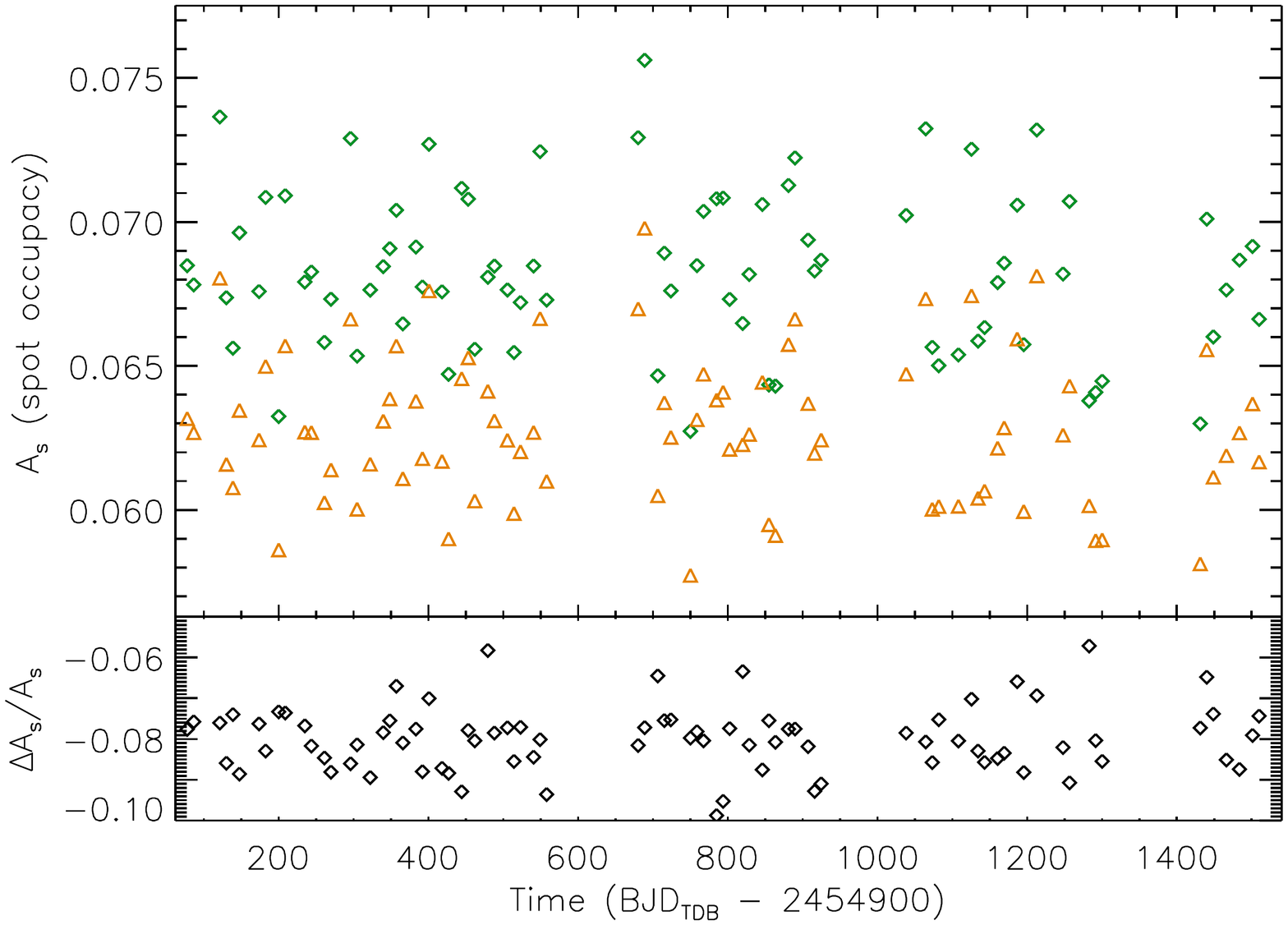}} 
\vspace*{-1.cm}
  \caption{Upper panel: Total coverage factor $A_{\rm s}$ of the starspots as derived from the ME models of the light curve of Bonomo \& Lanza with  facular-to-spotted area ratio $Q=2.4$ (green diamonds)  or $Q=1.0$ (orange triangles). Lower panel: relative difference between the values of the area obtained with the two different values of $Q$.}
              \label{area_comparison_Q1}%
\end{figure}
\begin{figure}
\hspace*{-1.cm}
 \centering{
 \includegraphics[width=8cm,height=10cm,angle=90]{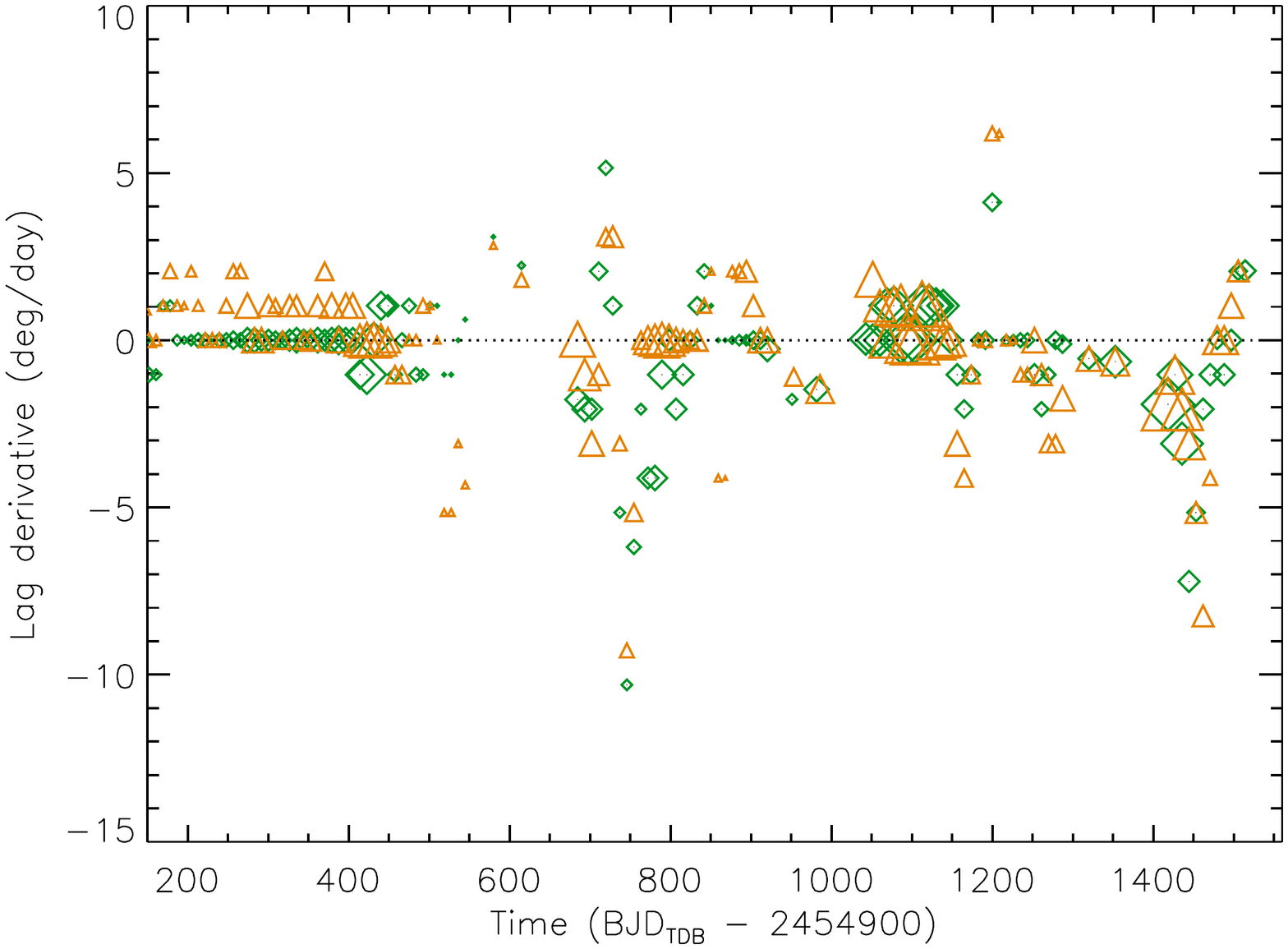}} 
\vspace*{-1.cm}
  \caption{Migration rate between consecutive spot pattern distributions as derived from the ME models of the Bonomo \& Lanza light curve  with facular-to-spotted area ratio $Q=2.4$  (green diamonds) or  $Q=4.0$ (orange triangles; see our Sect.~\ref{parameters}). The size of the symbols is proportional to the cross-correlation coefficient $\rho_{\rm cc}$ (cf. Eq.~\ref{cc_eq}). }
              \label{lag_migration_Q4}%
\end{figure}
\begin{figure}
\hspace*{-1.cm}
 \centering{
 \includegraphics[width=8cm,height=10cm,angle=90]{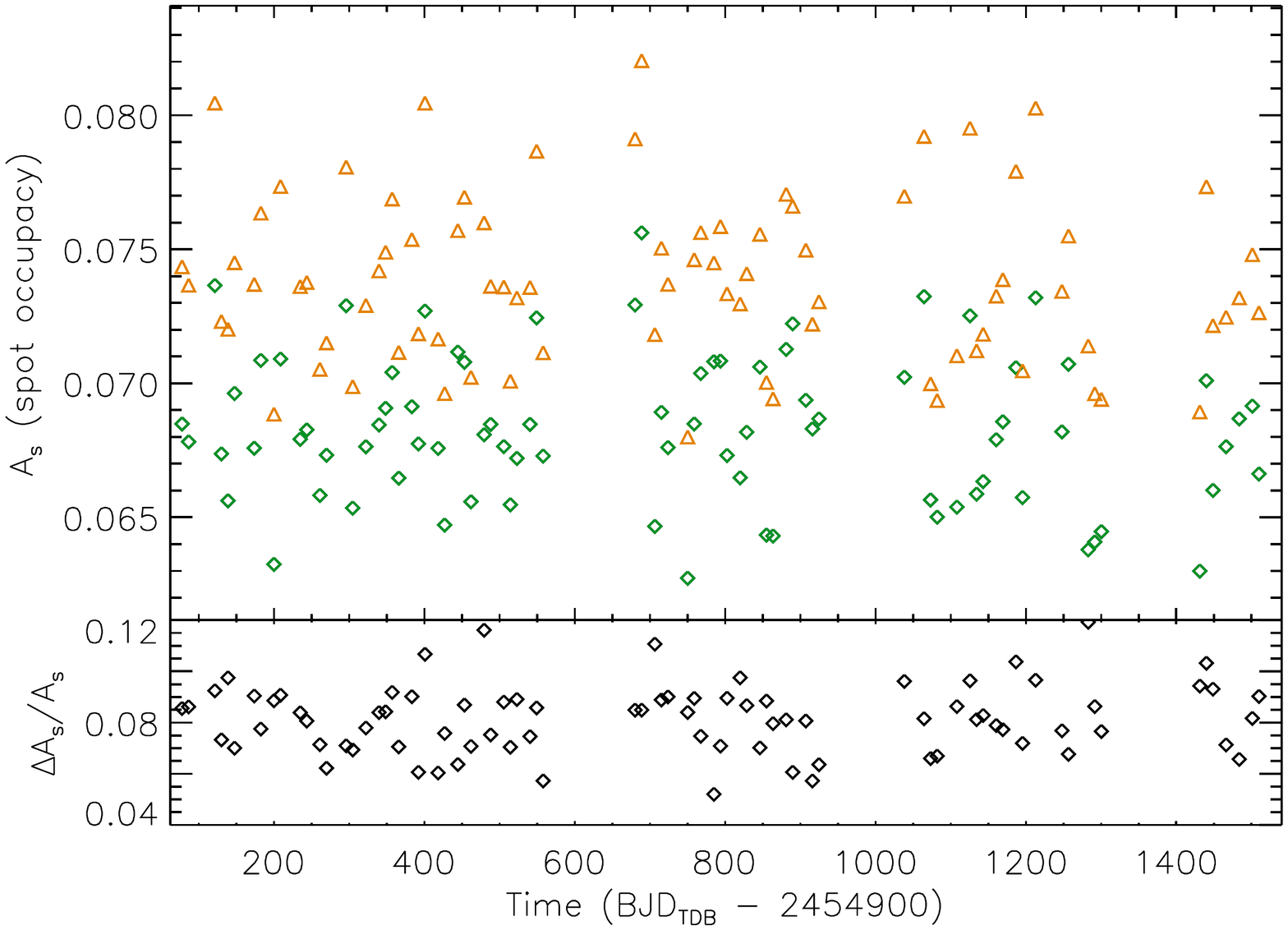}} 
\vspace*{-1.cm}
  \caption{Upper panel: Total coverage factor $A_{\rm s}$ of the starspots as derived from the ME models of the light curve of Bonomo \& Lanza with  facular-to-spotted area ratio $Q=2.4$ (green diamonds)  or $Q=4.0$ (orange triangles). Lower panel: relative difference between the values of the area obtained with the two different values of $Q$.}
              \label{area_comparison_Q4}%
\end{figure}
\section{A detailed view of the filling factor map}
\label{app2}
In Fig.~\ref{fig8_detail}, we show an enlargement of Fig.~\ref{fig8}. It shows the changes occurred between $t^{\prime} \simeq 750$ and $t^{\prime} \simeq 850$ when the spot pattern displayed an overall backward migration produced by a slower rotation, that is, a rotation period longer than $12.01$~days, the period of the reference frame adopted to plot the spot map. This backward migration was clearly detected by cross-correlating successive distributions of the filling factors as obtained from the ME models of the out-of-transit light curve (see Fig.~\ref{lag_migration}). However, we see in Fig.~\ref{fig8_detail} that individual longitudes show different migration rates also outside $750 \la t^{\prime} \la 850$~days indicating that they are produced by spots at different latitudes. Therefore, the migration rate given by the cross-correlation is an average over the whole longitudinal distributions. For $t^{\prime} \la 850$, the correspondence between the spots as mapped by the out-of-transit light curve and those mapped from transit occultations is poor (the cross-correlation at zero lag $\rho_{\rm cc}(0) \la 0.15$ in Fig.~\ref{correlation_in_out_transits})  suggesting that most of the former are located outside the occulted belt.  Individual spots are short-lived with typical lifetimes of a few tens of days, while active longitudes where spots form and decay are long-lived -- see the case of the active longitude around $200^{\circ}-250^{\circ}$ that disappears at $t^{\prime} \approx 770$ and re-appears for $t^{\prime} \ga 850$ days (see also Sect.~\ref{spot_active_longitudes}).  
\begin{figure*}
\hspace*{-1.cm}
 \centering{
 \includegraphics[width=18cm,height=21cm,angle=0]{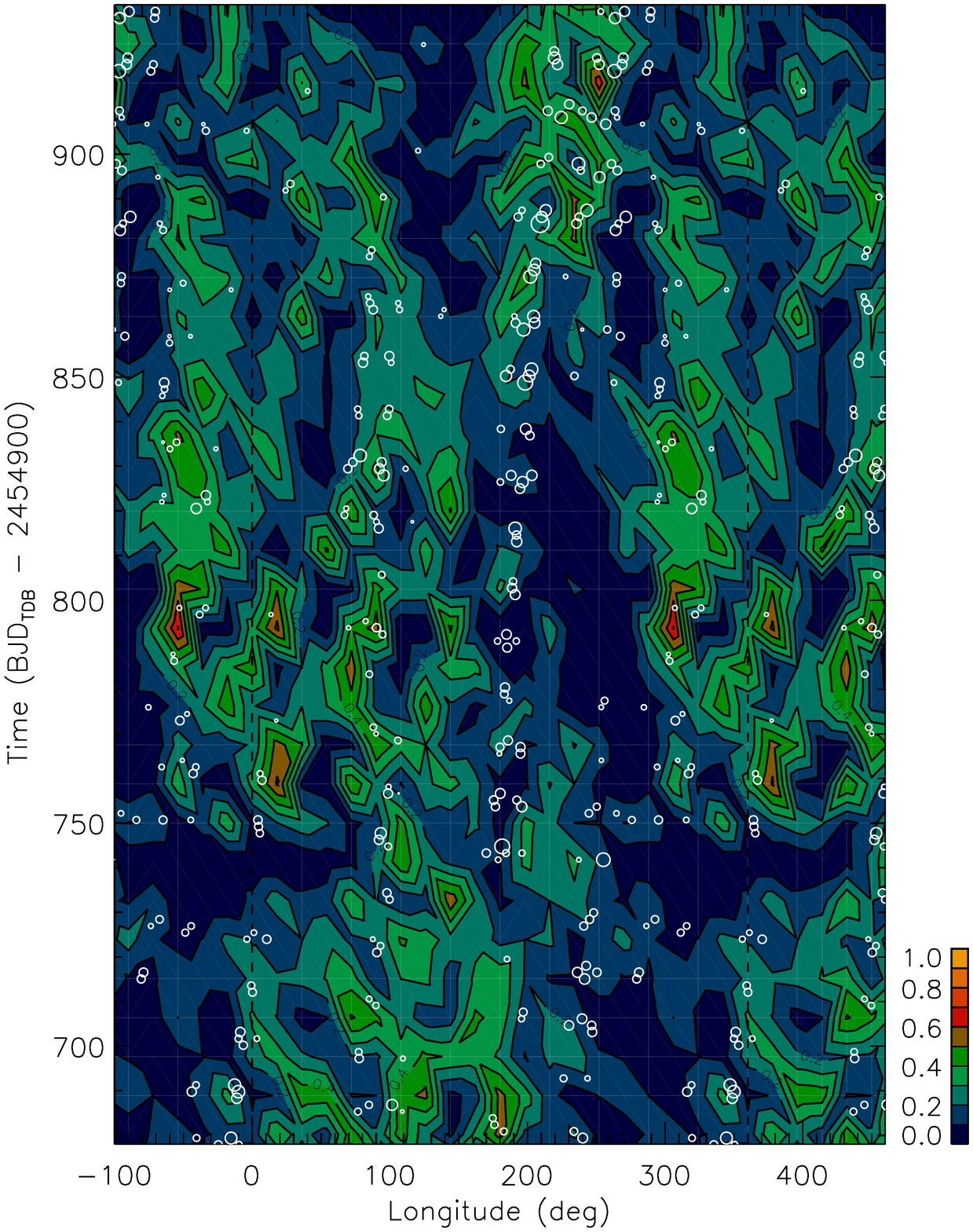}} 
\vspace*{-1.cm}
  \caption{An enlargement of Fig.~\ref{fig8} showing the distribution of the spot filling factor (see the colour scale at the bottom right) vs. the longitude and time as obtained by the ME modelling of the light curve de-trended as in \citet{BonomoLanza12}. The  spots detected during transits by \citet{Valioetal17} are overplotted as white circles the radius of which is proportional to their flux deficit $D$ as defined in Sect.~\ref{spot_active_longitudes}. The longitude scale goes beyond the interval $[0^{\circ}, 360^{\circ}]$ to help us following the migration of the spots. }
              \label{fig8_detail}%
\end{figure*}

\end{document}